\documentclass[aps,pre,twocolumn,superscriptaddress]{revtex4}
\usepackage{graphicx,amssymb,amsfonts,amsmath,chemarr,color,commath,braket,hyperref}

\begin{document}

\title{Temporal signals drive the emergence of multicellular information networks}

\author{Guanyu Li}
\thanks{These authors contributed equally to this work.}
\affiliation{Department of Physics, Oregon State University, Corvallis, OR 97331}

\author{Ryan LeFebre}
\thanks{These authors contributed equally to this work.}
\affiliation{Department of Physics and Astronomy, University of Pittsburgh, Pittsburgh, PA 15260}

\author{Alia Starman}
\affiliation{Department of Biomedical Sciences, Carlson College of Veterinary Medicine, Oregon State University, Corvallis, OR 97331}

\author{Patrick Chappell}
\email{Patrick.Chappell@oregonstate.edu}
\affiliation{Department of Biomedical Sciences, Carlson College of Veterinary Medicine, Oregon State University, Corvallis, OR 97331}

\author{Andrew Mugler}
\email{andrew.mugler@pitt.edu}
\affiliation{Department of Physics and Astronomy, University of Pittsburgh, Pittsburgh, PA 15260}

\author{Bo Sun}
\email{sunb@oregonstate.edu}
\affiliation{Department of Physics, Oregon State University, Corvallis, OR 97331}

\begin{abstract}
Coordinated responses to environmental stimuli are critical for multicellular organisms. To overcome the obstacles of cell-to-cell heterogeneity and noisy signaling dynamics within individual cells, cells must effectively exchange information with peers. However, the dynamics and mechanisms of collective information transfer driven by external signals is poorly understood. Here we investigate the calcium dynamics of neuronal cells that form confluent monolayers and respond to cyclic ATP stimuli in microfluidic devices. Using Granger inference to reconstruct the underlying causal relations between the cells, we find that the cells self-organize into spatially decentralized and temporally stationary networks to support information transfer via gap junction channels. The connectivity of the causal networks depend on the temporal profile of the external stimuli, where short periods, or long periods with small duty fractions, lead to reduced connectivity and fractured network topology. We build a theoretical model based on communicating excitable units that reproduces our observations. The model further predicts that connectivity of the causal network is maximal at an optimal communication strength, which is confirmed by the experiments. Together, our results show that information transfer between neuronal cells is externally regulated by the temporal profile of the stimuli, and internally regulated by cell-cell communication. 
\end{abstract}

\maketitle

\section*{Significance Statement}
Understanding how a group of cells cooperatively processes an environmental signal is an essential step to decoding the organizing principles of multicellular organisms. Here we demonstrate that neuronal cells form information-bearing causal networks through gap-junction-mediated communication. Our experimental and theoretical results uncover the mechanism by which excitable cells self-organize in response to external stimuli, and the rich collective dynamics enabled by non-synaptic intercellular interactions. Decoding such systems will lead to a better understanding of a diverse range of physiological processes, offering new insights into disease mechanisms and treatment.

\section*{Introduction}
Sensing and responding to chemical signals is of fundamental importance to living systems. For single cells, chemosensing is achieved by specialized receptors, which recognize molecules (ligands) in the microenvironment of cells \cite{Lahiri1993,Walter2002}. Such interactions trigger a cascade of intracellular events, which regulate the functional responses of cells, such as motility \cite{Voituriez&Ladoux_2021_migrationbyfootprint}, differentiation \cite{Basson_2012_differentiation&morphogenesis}, and gene expression \cite{Basson_2012_differentiation&morphogenesis, Wang_2008_geneexpression}. However, many chemosensing architectures determine the ligand concentration from time-integrated information, such as receptor-ligand binding and dissociation times \cite{Setayeshgar2005_physicallimits,Wingreen2005_MLE}. As such, dynamic external stimuli can present a challenge to cell sensing. For instance, oscillatory stimuli may be misinterpreted by cells, as external and internal time scales interfere in the signaling dynamics \cite{Lim2015_MAPK_confusing}.

In multicellular organisms, chemosensing is rarely accomplished by isolated single cells. Instead, collective chemosensing by communicating cells leads to rich dynamics that may be necessary to encode complex information \cite{ellison2016cell}. In collective chemosensing, environmental signals can induce specific single-cell dynamics that are regulated by cell-cell communication \cite{Morita_2017_calciumwavebycommunication}. For instance, we and other groups have shown that when chemosensing pathways support bifurcating signaling dynamics, cell-cell communication can shift the bifurcation boundary, so that the resulting cell response reflects both the external signal as well as the degree of communication \cite{Sun2013critical,Sun2016defective}.

Collective chemosensing can be manifested as orchestrated multicellular dynamics, such as intercellular synchronization. Synchronized cellular dynamics have been observed in cardiac tissues \cite{Agladze_2017_cardiacsynchronization}, endothelium \cite{Gerhardt_2016_endothelialsynchronization}, and in the hypothalamic suprachiasmatic nucleus \cite{Kramer_2007_SCNsynchronization}. Synchronization often requires strong external stimuli and efficient cell-cell communication to offset the intrinsic and extrinsic noise in the dynamics of individual cells \cite{Showalter2018_coupling}. 

Alternatively, collective chemosensing may induce a group of communicating cells to self-organize into networks that support asymmetric interactions and directed information flow. In particular, environmental stimuli facilitate the emergence of leader, follower, and pacemaker cells such as in beating cardiac tissues \cite{Moorman2010}, in social amoebae that form fruiting bodies \cite{gregor2010onset}, and in the neuronal regulation of circadian rhythms \cite{herzog2007neurons}. However, the hierarchical organization is often obscured by fluctuations of single cell dynamics, and requires sophisticated data analysis to reconstruct the underlying network. Information-theoretic metrics, such as Granger inference \cite{Granger_1969_grangercausality} and its non-parametric form of transfer entropy \cite{Barnett_2009_transferentropy} have been instrumental in elucidating the intercellular wiring hidden from direct observations \cite{Feng_2010_Granger}. Despite its biological significance, the underlying mechanisms, upstream control, and downstream function of collective chemosensing are still far from fully understood.\\

In this study we combine quantitative experiments and computational modeling of excitable cells to investigate the emergence of information-bearing networks when monolayers of neuronal cells sense extracellular ATP (Adenosine triphosphate). We examine the calcium dynamics of KTaR cells, a neuronal cell line we derived from KNDy (Kisspeptin, neurokinin B, and dynorphin) neurons within the arcuate nucleus of an adult female mouse \cite{Chappel_2016_Ktar-1}. We show that under periodic stimuli a group of interacting cells forms a directed causal network which maintains dynamic equilibrium over consecutive cycles of stimuli. The network characteristics not only depend on the level of communication between cells, but also on the temporal profile of the external driving. Together, we demonstrate that temporal signals from the environment instruct the self-organization and communication dynamics of a multicellular system.

\section*{Results}
\begin{figure}
\centering
\includegraphics[width=0.99\columnwidth]{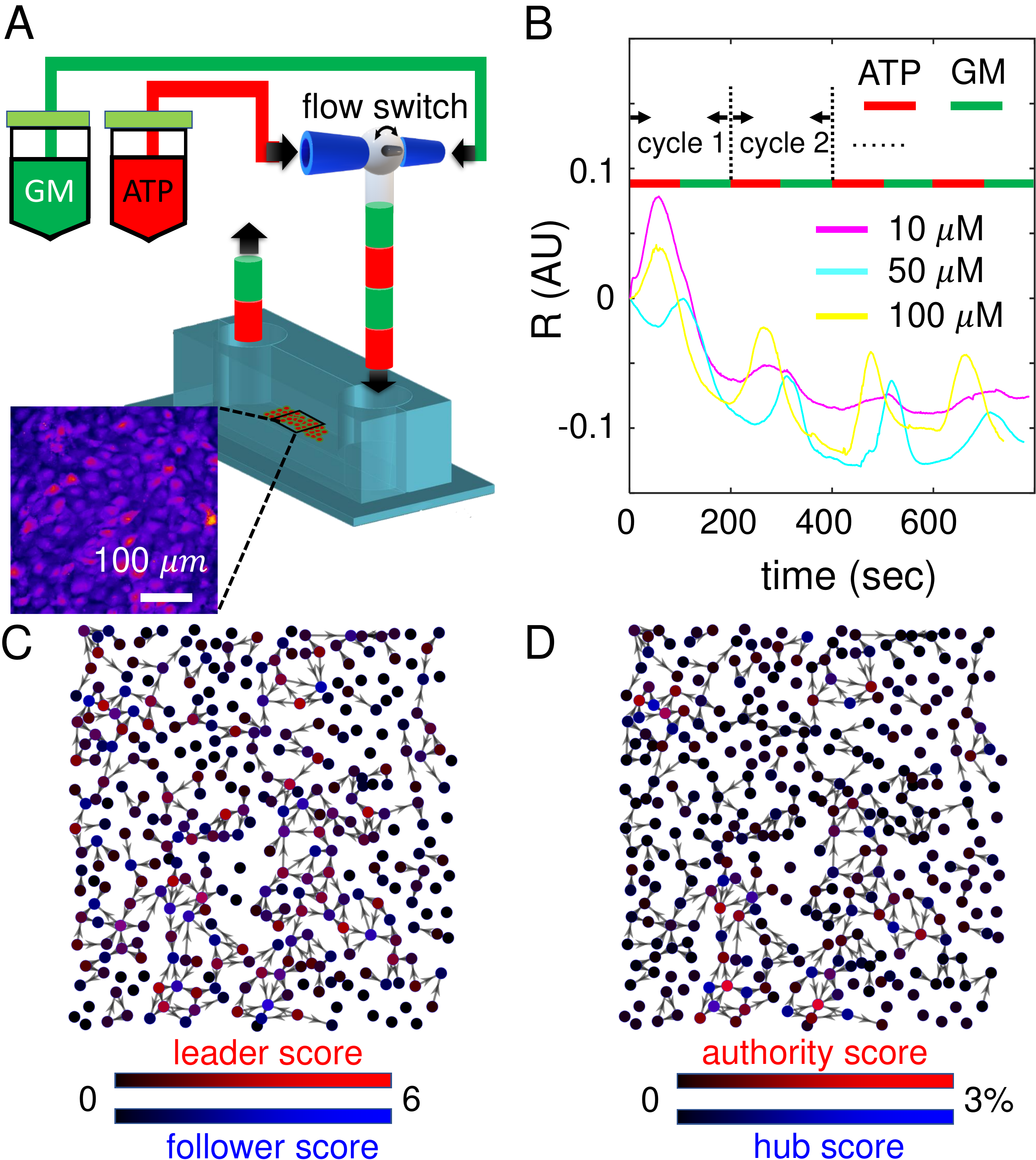}
\caption{Experimental setup to uncover the underlying self-organization of KTaR cell monolayers. (A) A schematic showing the microfluidics device to deliver alternate growth medium (GM) and ATP solution (ATP) to a confluent monolayer of KTaR cells. Inset: a fluorescent Ca $^{2+}$ image of a monolayer of KTaR cells. (B) Typical calcium responses $R(t)$ of KTaR cells to cyclic ATP stimuli at a period of 200 seconds. Heterogeneity among the cells lead to fluctuations in the magnitude and temporal delay of the calcium dynamics. (C-D) An example of reconstructed multicellular network via Granger inference. Direction of arrows point from a causing cell to its effected cells. Each node corresponds to the location of a cell in the field of view. The node are compositely colored by leader (red channel) and follower (blue) scores in C and authority and hub scores in D. }
\label{fig:fig1}
\end{figure}

In order to understand the collective dynamics of communicating cells under periodic stimuli, we employ a microfluidic device as shown in Fig.\ \ref{fig:fig1}A. A computer-interfaced flow switch alternates growth medium and ATP solution into the cell culture chamber, where a confluent monolayer of KTaR cells sense the ATP stimuli (see SI section S1a-c for more details of device and cell characterization). To detect cellular response, we preload the cells with a calcium indicator (Calbryte, AAT Bioquest), and record the fluorescent calcium images at single-cell resolution at 1 Hz for over 15 minutes. 

KTaR cells recognize extracellular ATP with purigenic receptors, which trigger IP3-mediated release of Ca$^{2+}$ from endoplasmic reticulum (ER) stores into the cytoplasm, as well as calcium influx from extracellular space \cite{Isakson_2014_purinergicreceptor} (SI section S1c-d).  While overall the relative change of intensity [$R_i(t)$, where $i$ is the cell index] follows the temporal profile of ATP stimuli, individual cells show variable phase delays to the global driving signal (Fig.\ \ref{fig:fig1}B, see also SI section S2). 

To quantify if the asynchronous responses of individual cells encode information transfer, we employ Granger inference \cite{Belliveau_2014_grangerdifference,Barnett_2015_grangerinneuroscience} to construct a directed graph that represents the causal influence between cells. Qualitatively, Granger inference designates a time series $C$ as causing a second time series $E$ if the combined history of both $\{C, E\}$ is significantly more predictive of time series $E$ than $E$'s own history alone. Because the rapid flow effectively washes away secreted factors \cite{Sun2012collective,Sun2016defective}, and because KTaR cells do not grow extended axons in our culture condition, we focus on nearest neighbor cells where gap junctional communication is dominant \cite{Kohn_2009_gapjunctionbetweenadjoiningcells} (SI section S1c). 

As we have shown previously, the time-derivative of fluorescent calcium intensity $\dot{R}_i(t)$ has the benefit of being independent of the basal intensity while still measuring communication effects \cite{Sun2016defective,Sun2013critical}. We have further confirmed that $\{\dot{R}_i(t)\}$ are stationary time series (SI section S2a), and therefore suitable for the application of Granger inference \cite{Barnett_2015_grangerinneuroscience}. 

For each nearest-neighbor pair, we calculate the statistical significance of Granger difference \cite{Belliveau_2014_grangerdifference} using the time series from a particular cycle. If higher than a threshold (95$\%$ confidence), an edge from the causal cell to the affected cell is drawn (see SI section S2b for more details). Fig.\ \ref{fig:fig1}C and D show an example of a reconstructed causal network, where each node represents the location of a cell in the field of view, and the arrows show direction of causality. 

After reconstructing the directed graph,  we have calculated the leader scores (number of outgoing edges) and follower scores (number of incoming edges) for each cell. The leader/follower scores distribute randomly in space (Fig.\ \ref{fig:fig1}C), indicating the absence of centralized organization. Indeed, we find the nodes generally have very low authority and hub scores as measured by Kleinberg Centrality \cite{Kleinberg1999} (Fig.\ \ref{fig:fig1}D), and the networks come with small Estrada index \cite{Estrada_2010_estradaindex} ($\lesssim$ 0.1, SI section S2c). These observations suggest that cyclic external stimuli trigger information transfer between communicating KTaR cells. Although heterogeneity among the cells prevents fully synchronized responses, the cells are able to self-organize into a decentralized causal network.

\begin{figure}
\centering
\includegraphics[width=0.99\columnwidth]{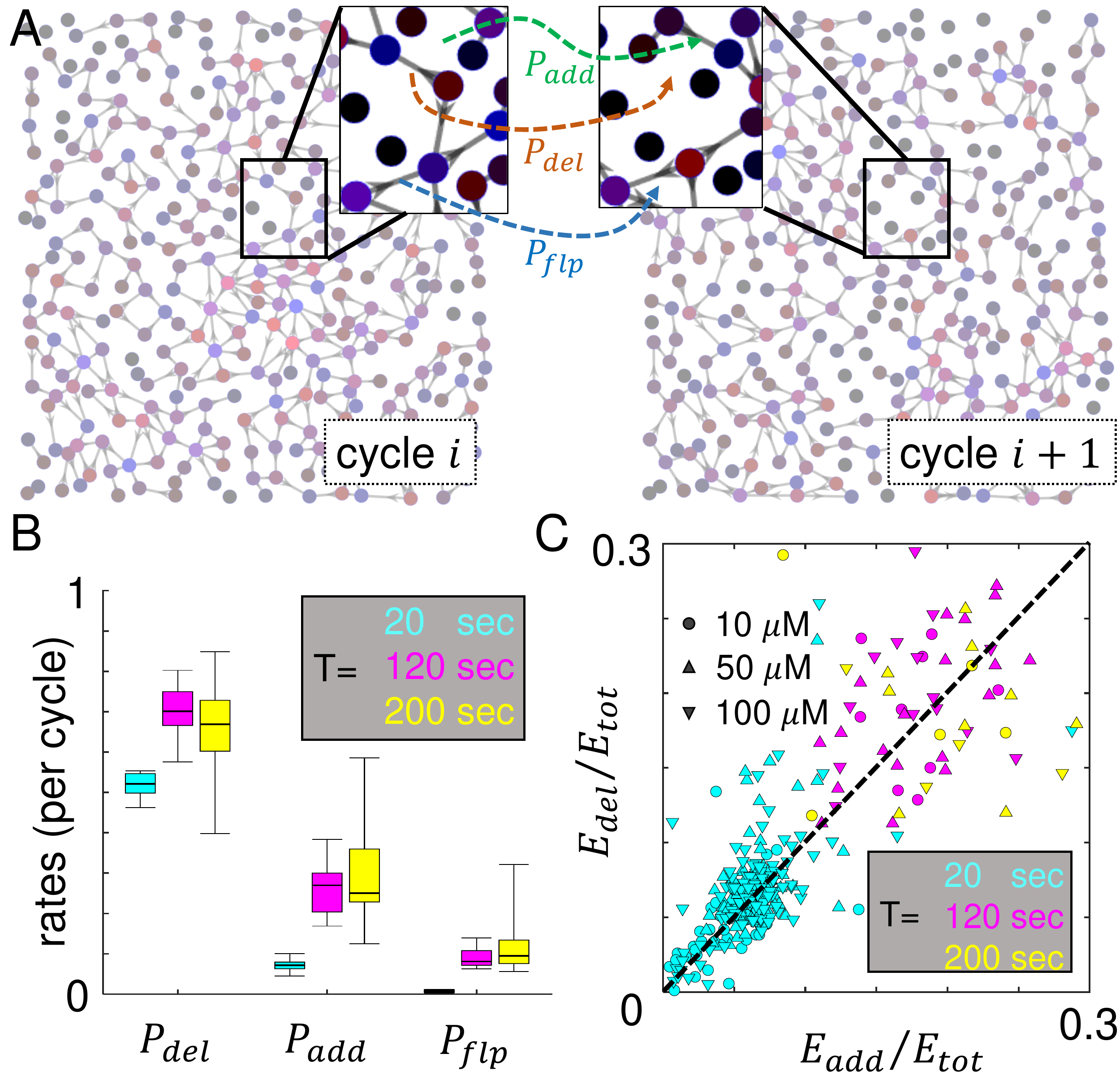}
\caption{Dynamic evolution of the multicellular networks driven by cyclic ATP stimulation. (A) An example multicellular network rewires between two consecutive cycles. The cells are exposed to 50 $\mu$M ATP at a period of 120 seconds. Insets show three types of rewiring events governed by their respective rates (probabilities per cycle): removing an edge ($P_{del}$), adding a new edge ($P_{add}$), and flipping the direction of an existing edge ($P_{flp}$). (B) The rates for removing, adding and flipping an edge at various driving periods $T$. Cyan: $T=$ 20 sec. Magenta: $T=$ 120 sec. Yellow: $T=$ 200 sec. For a given period, the rates do not depend on ATP concentration (see SI section S3b). (C) Scatter plot showing the numbers of added ($E_{add}$), and removed ($E_{del}$) edges between consecutive cycles normalized by the total number of nearest neighbors ($E_{tot}$). Here colors represent the driving period as in B. Different symbols represent ATP concentration. $\bullet$: 10 $\mu$M. $\blacktriangle$: 50 $\mu$M. $\blacktriangledown$: 100 $\mu$M.}
\label{fig:fig2}
\end{figure}

Having established methods to reconstruct the underlying networks of cells performing collective chemosensing, we first examine the evolution of the network structure over consecutive cycles of ATP stimuli (see also SI section S3a). To this end, we compute $P_{add}$, the rate (probability per cycle) of adding a new edge; $P_{del}$, the rate of deleting an existing edge; and $P_{flp}$, the rate of flipping the direction of an existing edge (Fig.\ \ref{fig:fig2}A).  We find approximately 60\% of edges are deleted from one stimulus cycle to the next, while a new edge would emerge from approximately 30\% of the unconnected neighbor cell pairs (Fig.\ \ref{fig:fig2}B). Among existing edges, less than 10\% of them will flip direction in the next cycle, indicating a memory effect that stabilizes the causal relation between cell pairs.

Although the values of $\{P_{add}, P_{del}, P_{flp}\}$ do not depend on the concentrations of ATP (SI, section S3b), nor the local connectivity (SI, section S3c), we find that at a period of 20 seconds all three rates are smaller compared with larger periods (Fig.\ \ref{fig:fig2}B). In all conditions, the network remains approximately stationary, as the number of new edges matches the number of removed edges over consecutive cycles (Fig.\ \ref{fig:fig2}C, see also SI section S3d). Taken together, these observations show that cyclic ATP stimuli drive monolayers of KTaR cells into networks that maintain their dynamic equilibrium.

\begin{figure*}
\centering
\includegraphics[width=0.99\textwidth]{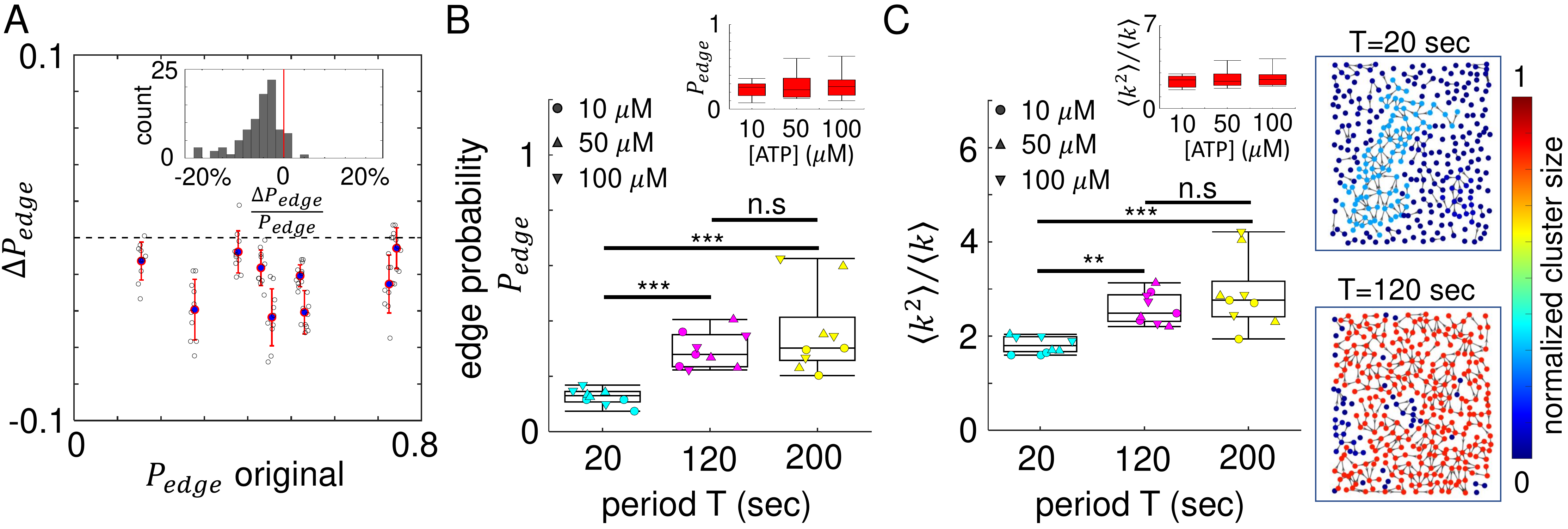}
\caption{Multicellular network connectivity is regulated by the period of ATP stimulation. (A) The edge probability of original experimental data and the change of edge probability $\Delta P_{edge}$ after randomization. In a particular experiment the randomization is done by switching the calcium dynamics of 10\% of randomly selected cell pairs in the same field of view. Inset: histogram of the relative change of edge probability after randomization. (B) The dependence of edge probability with respect to driving period. Inset: the dependence of edge probability with respect to ATP concentration. (C) The dependence of percolation degree $\langle k^2\rangle/\langle k \rangle$ with respect to driving period. Here $k$ represents the degree of the (undirected) network. Inset (top): the dependence of percolation strength with respect to ATP concentration. Inset (right): maps of clusters in two typical experiments ($T=$ 20 sec and $T=$ 120 sec). Nodes are colored by the size of clusters they belong to. Here the normalized cluster size is defined as the ratio between the number of cells in the cluster to the total number of cells in the field of view. In B-C colors of symbols represent the driving period. Cyan: $T=$ 20 sec. Magenta: $T=$ 120 sec. Yellow: $T=$ 200 sec. The types of symbols represent ATP concentration $\bullet$: 10 $\mu$M. $\blacktriangle$: 50 $\mu$M. $\blacktriangledown$: 100 $\mu$M. Statistical comparisons are done with ANOVA. **: $p<0.01$, ***: $p<0.001$, n.s. : not significant.}
\label{fig:fig3}
\end{figure*}

After showing the multicellular network to be stationary, we investigate whether the degree of network connectivity depends on the spatial relations between cells. We calculate the edge probability $P_{edge}$, which is defined as the number of edges divided by the number of nearest neighbors. In particular, we compare the edge probability of the original networks (directly obtained from experiments), and ones obtained by randomizing the original networks. The randomization is done by shuffling the time series of 10\% of the cells with another 10\% cells in the same experiment: $R_i\Longleftrightarrow R_j$, where $i\neq j$ are randomly chosen pairs. The randomized data encodes identical driving signal to the original data. We find that for all experiments even a 10\% partial randomization significantly dilutes the edges (Fig.\ \ref{fig:fig3}A), and $P_{edge}$ can be reduced by as much as 20\% (Fig.\ \ref{fig:fig3}A inset). The result highlights the locality of cell-cell interaction, which is consistent with gap-junction mediated information flow between nearby cells. 

Having demonstrated that the short-range intercellular communication is manifested by the edge probability, we now examine what aspects of the external signal control the network connectivity. To this end we systematically vary the ATP concentration $[$ATP$]$ and period of stimuli $T$. We find $P_{edge}$ increases dramatically when the driving period increases from 20 seconds to 120 seconds, and plateaus without further changing when the driving period is further increased to 200 seconds  (Fig.\ \ref{fig:fig3}B). At long driving periods, the edge probability falls in the range of 0.3 to 0.5, corresponding to 2-3 edges per cell (on average each cell has six nearest neighbors). On the other hand, the network characteristics do not depend on the ATP concentration over the physiological range of 10 $\mu$M to 100 $\mu$M \cite{Buffington_2007_HumanATPconcentration} (Fig.\ \ref{fig:fig3}B inset). 

To further compare the self-organized structure of multicellular networks at varying driving signals, we compute the percolation degree $N = \langle k^2\rangle/\langle k\rangle$. $N$ is the average degree (i.e., total number of edges to a node) of a node if the node is linked to another node. When $N>2$, the network is above the percolation threshold and features large connected components (clusters). When $N<2$, the network is below the percolation threshold and consists of many small clusters \cite{Havlin2010}. We find when the driving period equals 20 seconds, the percolation degree is less than 2 (Fig.\ \ref{fig:fig3}C). A typical network in this case (Fig.\ \ref{fig:fig3}C top right inset) indeed shows fractured topology where none of the clusters contains more than 20\% of the cells in the field of view.  In contrast, at larger driving period the percolation degree is greater than 2 (Fig.\ \ref{fig:fig3}C) such that a typical network is dominated by a single large cluster (Fig.\ \ref{fig:fig3}C bottom right inset). These results show that there exists a critical time scale of the external driving signal that dictates the underlying information flow of collective sensing. At a small driving period, KTaR cells form a loosely connected, fractured network. Conversely at large driving periods, highly connected and percolating networks emerge thanks to the elevated information flow between cells. These two distinct types of multicellular organization are induced by the temporal profiles of the driving signal, rather than the concentration of the stimuli.

\begin{figure}
\centering
\includegraphics[width=0.99\columnwidth]{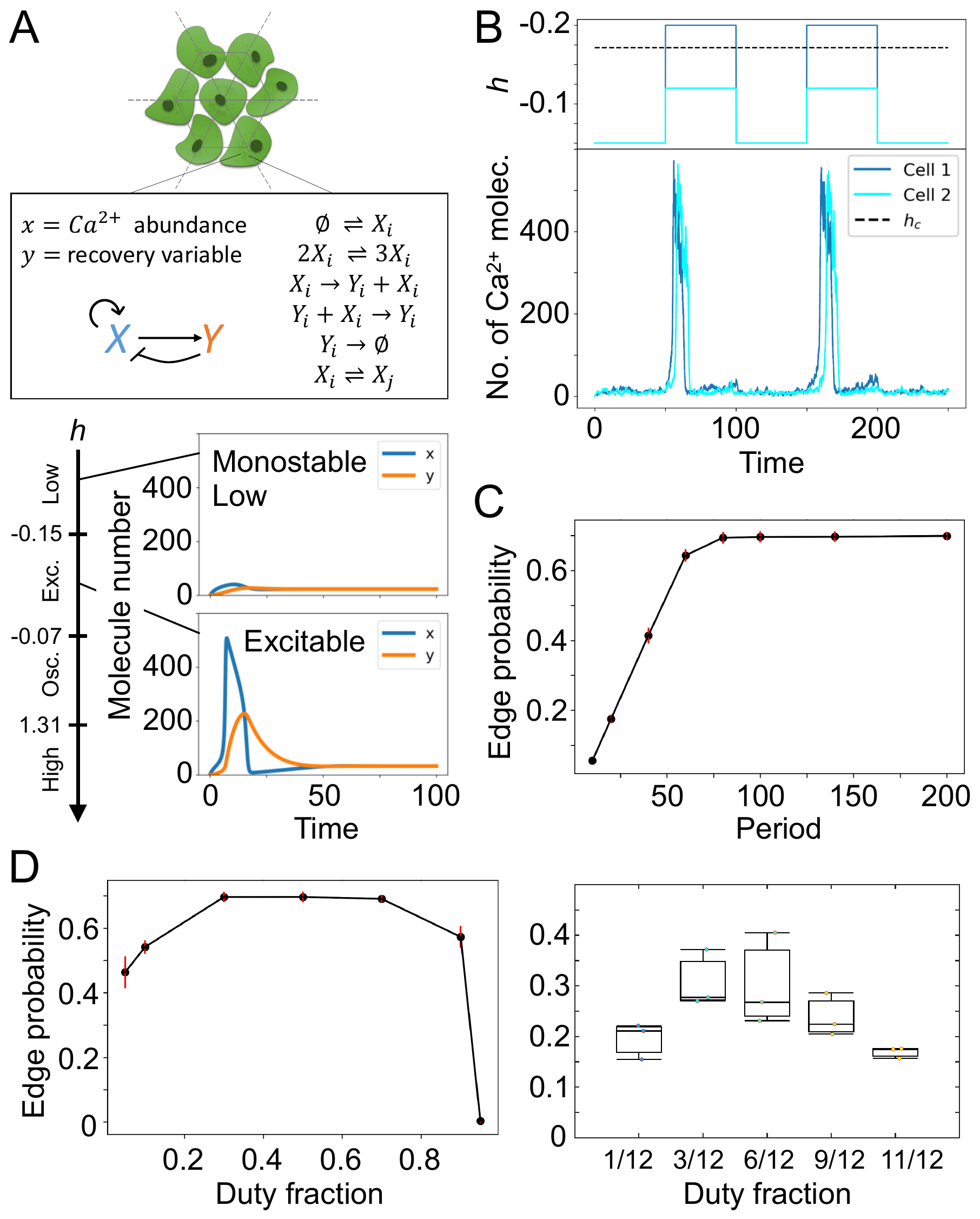}
\caption{Mathematical model of collective excitable dynamics. (A) Each cell in a triangular lattice contains a calcium variable subject to positive and negative feedback reactions (via a slower recovery variable) to enable excitations, and exchange reactions to enable nearest-neighbor communication. (B) An excitable cell (Cell 1, $h>h_c$) can induce a non-excitable cell (Cell 2, $h>h_c$) to excite with a delay via cell-cell communication. (C) Fraction of nearest neighbors with causal edges (edge probability) increases and saturates with stimulus period in model. Here the presence of an edge is determined by the delay between neighbors' excitations (see SI section S4). (D) Edge probability decreases for small or large duty fraction (fraction of period for which stimulus is on) in model (left) and experiments (right). Experiments are done with 50 $\mu$M ATP at a period of 120 seconds. In D and E (left), error bars are standard deviation over 100 simulations of 8 cycles.}
\label{fig:fig4}
\end{figure}

To understand the mechanisms by which temporal signals drive the emergence of multicellular networks, we developed a mathematical model of communicating excitable cells. Each cell is modeled using a reduced form of the Hodgkin-Huxley model \cite{hodgkin1952quantitative}, which is widely accepted to replicate neuronal dynamics. Specifically, within each cell on a six-neighbor triangular lattice (Fig.\ \ref{fig:fig4}A), two chemical species interact, which is the minimum needed for excitable dynamics \cite{strogatz2018nonlinear}: $X$, which represents calcium abundance; and $Y$, which represents a slower recovery variable. The following minimal reactions are chosen to produce excitations: $X$ activates both itself \cite{schlogl1972chemical, erez2019universality} (Fig.\ \ref{fig:fig4}A, first two reactions) and $Y$ (third reaction), while $Y$ represses $X$ (fourth reaction). Both $X$ and $Y$ degrade spontaneously (first and fifth reaction), and $X$ is exchanged between neighboring cells to model the gap-junction communication (sixth reaction). Transforming the rate equations of this model into a standard form (see SI section S4) makes clear that the dynamics are specified by (i) a characteristic molecule number $x_c$, (ii) a timescale separation $\epsilon$ between $X$ and $Y$, and (iii) an external ``field'' $h$ that tunes the system among four regimes: stable dynamics at low molecule number, excitable dynamics, oscillatory dynamics, and stable dynamics at high molecule number (Fig.\ \ref{fig:fig4}A). The standard form is akin to the FitzHugh-Nagumo model \cite{fitzhugh1961impulses, nagumo1962active}, which is a reduced representation of more complex excitation models such as the Tang-Othmer \cite{tang1995frequency} and Hodgkin-Huxley \cite{hodgkin1952quantitative} models. Here we focus on the transition between the stable low and excitable regimes at $h=h_c$, and the effects of communication on this transition.

Modeling ATP as setting the value of the field $h$, we find that communication between an excitable cell ($h>h_c$) and a non-excitable cell ($h<h_c$) can induce an excitation in the non-excitable cell (Fig.\ \ref{fig:fig4}B), albeit with a delay. Indeed, applying Granger inference to stochastic simulations \cite{gillespie1977exact} of the model, we find that the excitable cell ``Granger-causes'' the non-excitable cell in this case. For simplicity, in larger networks we take the peak delay between neighboring cells as a proxy for the Granger metric when determining edges, as we find that the two are correlated in the simulations (see SI section S4). We then investigate networks of similar size to the experimental viewing window in which field strengths $h$ are drawn from a normal distribution centered just above $h_c$, such that slightly more than half of cells are excitable from the stimulus alone.

In these model networks we find that the edge probability increases and then saturates with the driving period (Fig.\ \ref{fig:fig4}C). This finding is consistent with the experiments (Fig.\ \ref{fig:fig3}B), which validates the model. The intuitive reason is that when the driving period is shorter than the excitation timescale, cells are still in the recovery phase and cannot respond, which reduces causal information. This intuition also holds when fixing the period but varying the duty fraction: if either the on- or off-portion of the cycle is too brief, the edge probability is reduced (Fig.\ \ref{fig:fig4}D, left). Varying the duty fraction in the experiments, we see that this prediction is upheld (Fig.\ \ref{fig:fig4}D, right), in further support of the model.

\begin{figure}
\centering
\includegraphics[width=0.99\columnwidth]{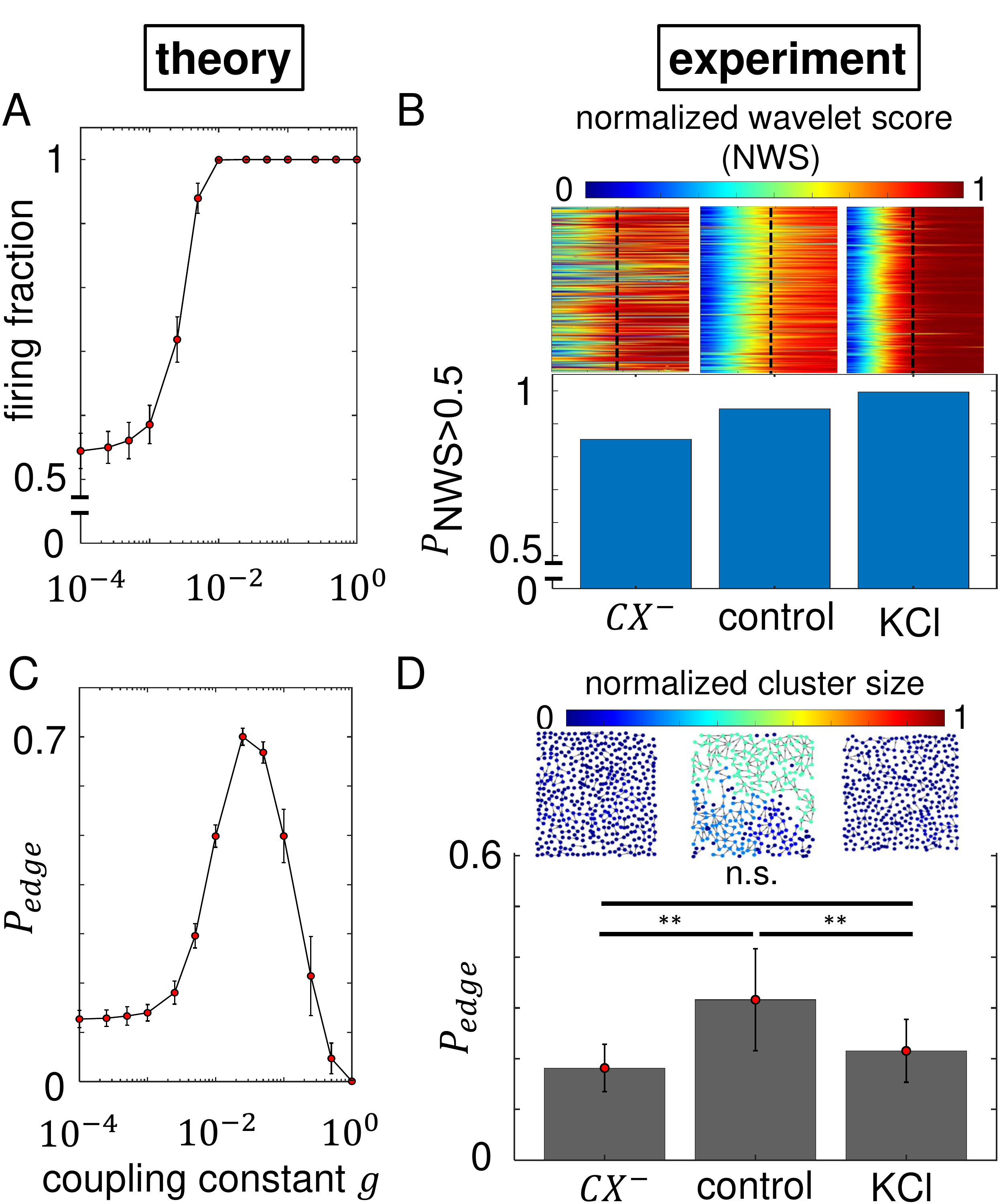}
\caption{Collective sensory responses at varying levels of communication. (A) Model prediction of the fraction of firing cells as the coupling constant $g$ changes over four decades. (B) The fraction of KTaR cells with normalized wavelet score (NWS) greater than 0.5 for three cases: cells are treated by 10 $\mu$M palmitoleic acid to inhibit gap junction and exposed to 50$\mu$M ATP at a period of 120 sec (CX$^-$); untreated cells are exposed to 50$\mu$M ATP at a period of 120 sec  (control); cells are exposed to 30 mM KCl at a period of 120 sec (KCl). Inset: the normalized wavelet scores (NWS) shown as heatmaps. Horizontal axis of the heatmaps represent time, where the black dashed lines indicate $t=2T=$ 240 seconds. NWS of individual cells are stacked vertically. See SI section S3 for more details. (C) Model prediction of the edge probability as the coupling constant $g$ changes over four decades. (D) Experimentally measured edge probability for the same conditions as in B. Statistical comparisons are done with ANOVA. **: $p<0.01$, ***: $p<0.001$, n.s. : not significant. $N=3$. Errorbars: standard deviation.}
\label{fig:fig5}
\end{figure}

Having investigated the relationship between the external driving and intracellular excitation timescales, we now use the model to investigate the effects of changing the strength of intercellular communication. Upon varying the cell-cell coupling constant $g$ over four decades, we find that stronger communication leads to a higher fraction of cells exhibiting an excitable response, $x>x_c$ (Fig.\ \ref{fig:fig5}A). Evidently, for sufficiently strong communication, all non-excitable cells can be induced to excitation by communication alone. This result is consistent with our previous report in fibroblast monolayers that communication can augment the effects of external stimuli by modulating the bifurcation threshold of excitable cell dynamics \cite{Sun2016defective}.

To test the model prediction, we devised methods to either reduce or enhance intercellular communication experimentally (SI section S1). To reduce communication, we treat the KTaR cells with 10 $\mu$M, palmitoleic acid, a broad spectrum gap junction inhibitor \cite{SALAMEH200536}. To enhance communication, we replace ATP with KCl as the external stimulus. KCl depolarizes the KTaR cell membrane, triggering an action potential as well as intracellular calcium responses \cite{Garcia2020_KCL}. As a result, the cells couple electrically through gap junctions, which is faster compared with the diffusion-limited molecular exchange.

Unlike in the model where the criterion for excitation is self-evident ($x>x_c$), in the experiments the criterion must be defined from the response itself. To this end, we define a normalized wavelet score (NWS) to quantify the cellular dynamics in the frequency domain using a time-resolved wavelet transformation (SI section S3). If the calcium dynamics of a cell perfectly follows the driving frequency, its NWS equals 1 at all times (except for boundary effects that affect the beginning and end of the time series). Otherwise, the NWS will fluctuate between 0 and 1 when irregular response occurs.

We find that the NWS of cells treated with palmitoleic acid (abbreviated as $CX^{-}$) show significantly stronger fluctuations compared with untreated cells, whereas the NWS of cells stimulated with KCl quickly reach and stay close to 1 (Fig.\ \ref{fig:fig5}B, top). To compare with the model prediction, we calculate the fraction of cells with an NWS greater than 0.5 at time point $t=2T$ to avoid boundary effects, where $T$ is the driving period set to be 120 seconds. Consistent with the model, we find the fraction of cells with NWS greater than 0.5 is highest for KCl excited cells, and lowest for gap junction inhibited cells (Fig.\ \ref{fig:fig5}B, bottom).

Interestingly, our model also predicts that the network connectivity reaches a maximum at an optimal coupling constant $g$, and decreases in either direction from the optimal value (Fig.\ \ref{fig:fig5}C). The intuitive reason is that with weak communication, only the inherently excitable cells are responding, such that causal edges do not form with non-excitable cells. Intermediate communication induces non-excitable cells to excite with a delay, introducing new edges. Strong communication synchronizes cells, reducing causality and removing edges.

Experiments confirm that the edge probabilities for gap junction inhibited monolayers, and for KCl excited monolayers are both lower than the untreated KTaR cells exposed to ATP stimuli (Fig.\ \ref{fig:fig5}D). Consistently, typical networks of untreated cells show characteristics of percolation, while networks under the other two conditions are evidently fractured (Fig.\ \ref{fig:fig5}D inset). These observations suggest that under the control condition the KTaR monolayers are posed close to the optimal coupling strength for causal information flow. Inhibiting gap junction curtails cell-cell communication, reducing network connectivity, whereas accelerating cell-cell communication leads to rapid synchronization between neighboring cells, also reducing information flow. Together, our results demonstrate that the self-organization of multicellular networks is modulated by the level of cell-cell communication.

\section*{Discussion}
A group of interacting cells encodes environmental information in different forms than single cells do. Revealing the underlying principles of collective chemosensing is an essential step to understanding the rules of life. Here we study the external ATP-triggered calcium dynamics of neuronal cell monolayers. We employ microfluidics to deliver alternate ATP solution and pure growth medium to KTaR-1 cells, a neuronal cell line we derived from KNDy neurons within the arcuate nucleus of an adult female mouse. KTaR cells express connexin proteins \textit{in vitro} which constitute gap junction channels between adjacent cells. Using Granger inference, we show that during each ATP-growth medium cycle, there is asymmetric information flow between adjacent cells manifested as causal relations between their intracellular calcium dynamics. As a result, the external stimuli drive the neuronal cell monolayers to establish directed networks. These networks display hierarchical structure where leader and follower cells distribute spatially without any apparent centralized organization (Fig.\ \ref{fig:fig1}).

The information networks are highly dynamic from one cycle of stimuli to the next, while the overall connectivity remains stationary. For all conditions tested, most structural fluctuations of the networks manisfest as adding or removing edges, whereas less than 10\% of the edges flip directions over consecutive cycles (Fig.\ \ref{fig:fig2}B). This suggests that the network reconfiguration is due to stochastic disappearance and reappearance of deterministic causal relationships that presumably arise from cell-to-cell heterogeneity \cite{Wollman2018}. The time evolution of the networks show characteristics of detailed balance. For instance, the number of edges remain approximately constant (Fig.\ \ref{fig:fig2}C). The probability flux in the cellular state space defined by the leader/follower scores also vanishes (SI section S2). This is in contrast to the nonequilibrium stationary states observed in other living systems, especially in the macroscopic brain dynamics \cite{Bassett_2021_brokendetailedbalanceinbrain}. It is conceivable that higher order organization in the brain leads to the emergence of entropy production that is absent at the scale of locally communicating neuronal cells. 

Many neuronal systems  demonstrate characteristics of learning and reinforcement \cite{Niv_2008_reinforcementlearninginbrain}. In contrast, we find that under repeated stimuli, gap junctions mediate a Markovian evolution of KTaR networks that keeps the system stationary. This is expected, as gap junctions alone have rapid turnover time \cite{Pereda2012_gapjunction_tracking}. It will be interesting for future studies to elucidate the mechanisms by which neuronal cells stabilize their information exchange dynamics.

We find that the edge probability of the multicellular network primarily depends on the timescale, and is impervious to the magnitude of external stimuli (Fig.\ \ref{fig:fig3}). Interestingly, both the experiments and the theoretical model show that the effective timescale of the external signal is determined by the lesser of the on- and off-duty cycles (Fig.\ \ref{fig:fig4}).
It makes sense that short on-times may be insufficient to trigger excitations (or sustain neighbor-induced excitations), but this result implies that short off-times are also insufficient to do so. This is likely a result of the need for a post-excitation recovery time, which is a generic property of excitable systems. Indeed, the minimal nature of the model suggests that our findings on network responses to temporal signals may be generalized to other multicellular excitable systems.

Our finding that an intermediate communication strength maximizes causal connectivity (Fig.\ \ref{fig:fig5}) has implications for information propagation in multicellular systems. In systems unlike ours, where a stimulus is localized or the medium itself is spatially directed, one expects that causal information should increase indefinitely with the communication strength between units. However, in systems like ours, where neither the stimulus nor the medium break symmetry, our results highlight an interesting regime where intermediate communication amplifies heterogeneity to create random but reproducible pathways of information flow. Such a regime may be important in systems where these initially spontaneous pathways are reinforced and built upon to break symmetry permanently, facilitating the formation of differentiated structures \cite{goryachev2021symmetry}.

\textit{In vivo}, KNDy neurons are critical for pubertal progression and sex steroid feedback involved in the neuroendocrine regulation of reproduction, expressing and secreting Kisspeptin, a peptide stimulatory to gonadotropin-releasing hormone (GnRH) secretion \cite{Lehman2018_KNDy}. It is also interesting for future investigations to determine the physiological effects of temporal sensitivity for KNDy neurons \textit{in vivo}, which receive episodic afferent signals including glutamate, GABA, neurokinin B in addition to ATP. 

Broadly speaking, collective chemosensing in a multicellular system may behave as isolated non-interacting units, may form hierarchical information flow networks, or may achieve synchronization. Here we show that the transition between these scenarios is controlled by the interplay of two timescales: one that is set internally by the communication channels between the cells, and one that is set externally by the driving signal. Our results highlight the rich dynamics exhibited by spatially coupled excitable units. Decoding such systems will lead to a better understanding of a diverse range of physiological processes from development to neuronal dynamics to tissue organization \cite{Shvartsman_2021_clonaldominance}, offering new insights into disease mechanisms and treatment \cite{Arnsdorf1991,KazimENEURO2020}.

\section*{Materials and Methods}
See the SI Appendix for details of cell culture, microscopy and image analysis. The statistical analysis and computer simulations are performed with Matlab (MathWorks\textregistered). See the SI Appendix for details of the theory and the simulations. 

\section*{Data Availability}
Experimental recordings of calcium dynamics are available at \url{https://figshare.com/s/a681463e1e69134f7320}. Additional microscopy images of the experiments are available by request to the corresponding authors. Simulation codes are available at \url{https://github.com/rwl23/mutlicellular_information_networks}.

\section*{Acknowledgments}
GL is supported by National Science Foundation PHY-1844627. RL, AM, GL, AS and PC are supported by National Institute of General Medical Studies grant R01GM140466. BS is supported by National Institute of General Medical Sciences grant R35GM138179.

\section*{Author Contributions}
BS and AM initiated the project. GL, AS, PC conducted the experiments. RL and AM developed the theoretical model. All authors analyzed data and wrote the manuscript.


\onecolumngrid
\section*{Supplemental Material}

\section*{S1. Additional information of experimental system}
\subsection*{a.	Fabrication and characteristics of microfluidic device }
\begin{figure}[h]
\centering
\includegraphics[width=0.8\textwidth]{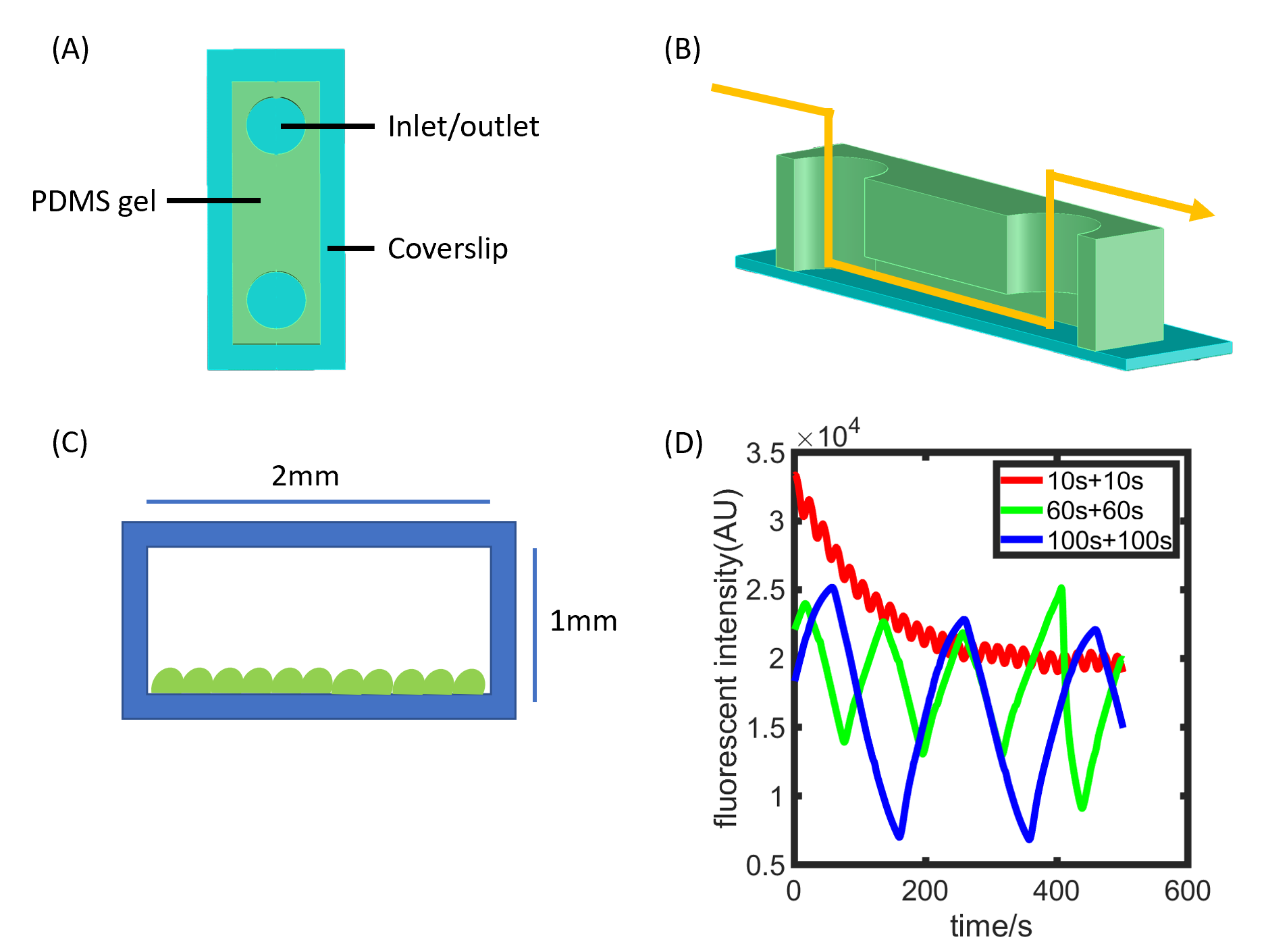}
\caption{Schematic of the PDMS device and fluorescent intensity data with different switching periods. (A) Top view of the PDMS device shows the inlet/outlet, the glass slide and the PDMS gel. (B) Side view of the PDMS device, and arrow indicates the direction of flow. (C) Cross section views of the channel shows the dimension of the channel and attached KTaR-1 cells. (D) Temporal profiles of brightness intensity inside the chamber while continuous perfusion was operated by fluorescein/water. All three different periods, $T=$20 sec (10s+10s, meaning 10 seconds of stimuli followed by 10 seconds of water/growth medium each cycle), $T=$120 sec (60s+60s) and $T=$200 sec (100s+100s) are tested. Image is recorded after flow is stabilized. }
\label{SIfig:PDMS}
\end{figure}

The polydimethylsiloxane (PDMS, Sylgard 184, Dow-Corning) that we use for making the PDMS device contains two parts, a low-matched base and a curing agent. We mix the base and curing agent as the manual instructs, then degas, and is poured over a stainless-steel mold before curing at 80 $^\circ$C for 8 hours to overnight. After the mixture is cured, the PDMS gel would be cut from the mold, and the inlet/outlet would be punctuated by a 1.5 mm OD probe needle. After that, the PDMS gel and No.1 coverslip would be corona treated and stuck together by adding heavy weight on top while heating on 200$^\circ$C for 4 to 8 hours. The following schematics demonstrate the design of the devices. Fig. \ \ref{SIfig:PDMS}A-C provides the schematic of the device as well as inner dimension of the device.
\bigbreak
\noindent The stimuli profile is characterized by recording fluorescent intensity of alternative perfusion between water and fluorescein (Sigma Aldrich, MO) using the same flow condition as in ATP experiments. Fig  \ \ref{SIfig:PDMS}D provides the fluorescent intensity under three different switching periods. The flow rate is set to be 130 $\mu$L which ensures minimal drifting of the cells.

\subsection*{b.	Cell Culture and Sample Preparation }
The cell line is cultured in standard growth medium (Dulbecco’s Modified Eagle Medium (DMEM, Sigma-Aldrich, MO) supplemented with 10\% Fetal Bovine Serum and 2\% 100x penicillin) in T-25 flasks. Cells will be subcultured and used for experiments when they reach 80\%-90\% confluency.
\bigbreak
\noindent To prepare the sample, cells are detached from culture flasks using TrypLE Select (Life Technologies, CA) for 10 mins at incubator under standard condition (37$^\circ$C, 5\% humidity), and then being centrifuged at 123g for 5 mins and being suspended in growth medium before pipet into the microfluidic devices. Finally, cells are injected into microfluidic device with desired density (~2000 cells/$mm^2$). The cells will be allowed to attach the bottom glass and constructs gap junction by incubating at 37$^\circ$C, 5\% humidity incubator for 8 to 12 hours. 
\bigbreak
\noindent On day 1 of experiments, fluorescent calcium indicator is prepared by using Calbryte™ 520 AM (AAT bioquest, CA) following the receipt below: 32 $\mu$L DMSO will be mixed with 50g of Calbryte™ 520 AM to make 1.4 mM solution. The final solution will be composed of 8 $\mu$L of 1.4 mM Calbryte™ 520 AM solution with 64 $\mu$L of 25 mM probenecid, 728 $\mu$L of HHBS and 800 $\mu$L of complete growth medium. Then, 100 $\mu$L of the final solution is added into the microfluidic device and incubated for 40 to 50 mins in the incubator. When performing multiple experiments, devices are labelled one by one to prevent the apoptosis due to toxicity of the dye. After taking the device out of incubator and adding fresh growth medium, the device is then ready for imaging.

\subsection*{c.	Characterization of KTaR cells }
The KTaR-1 cells (referred to as KTaR cells in the main text) used in this study are immortalized kisspeptin (Kiss-1) neurons derived from the arcuate nucleus of an adult female mouse\cite{Chappel_2016_Ktar-1}. These neurons are isolated from \textit{kiss1}-GFP mice, and immortalized by genomic integration of large and small T antigen, using a 3rd-generation lentiviral delivery system from Addgene. KTaR-1 neurons express kiss1, Neurokinin B (\textit{tac2}) and Dynorphin (\textit{pdyn}), making them appropriate models for investigation of properties of KNDy arcuate hypothalamic neurons, which are crucial for normal pubertal progression and fertility in females. Similar to what is observed \textit{in vivo}, these neurons express variable levels of \textit{kiss1} depending on exposure to estrogen (17$\beta$-estradiol), with low picomolar concentrations of this steroid hormone acting to repress \textit{kiss1} expression.  
\bigbreak
\noindent We have further confirmed common gap junction proteins are expressed in KTaR cells. In particular, as shown in Fig. \ref{SIfig:PCR} KTaR cells express connexin 43 (\textit{gja1}), connexin 37 (\textit{gja4}) and connexin 26 (\textit{gjb2}) proteins, as well as the inwardly-rectifying ATP-gated potassium channel Kir3.4 (\textit{kcnj5}). Expression of gap junctional hemichannels was compared in three murine hypothalamic neuronal lines (KTaR-1, KTaV-3, and GT1-7) and mouse brain as a positive control. NTC= no template negative control.

\begin{figure}[h]
\centering
\includegraphics[width=0.8\textwidth]{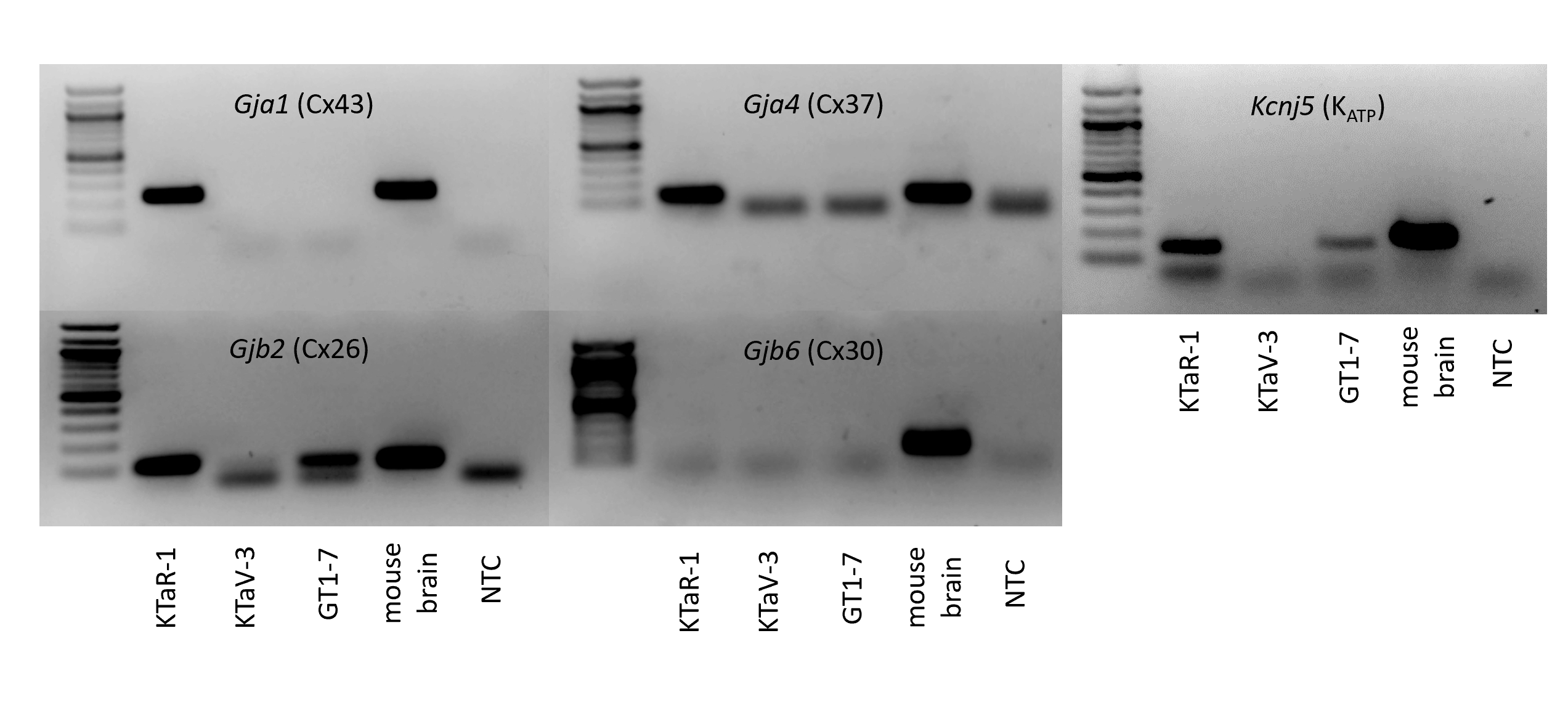}\label{SIfig:PCR}
\caption{Expression of gap junction hemichannel connexins in KTaR-1 cells, compared with other murine hypothalamic neuronal cell lines and total mouse brain examined with RT-PCR. KTaR-1 neurons express \textit{gja1} (Cx43; A), \textit{gjb2} (Cx26; B), and \textit{gja4} (Cx37; C). No expression of Cx30 was found in KTaR neurons (D). These cells also express the K$_{\text{ATP}}$ channel \textit{kcnj5} (E).}
\label{SIfig:PCR}
\end{figure}

\subsection*{d.	The molecular pathway of calcium responses}
\begin{figure}[h]
\centering
\includegraphics[width=\textwidth]{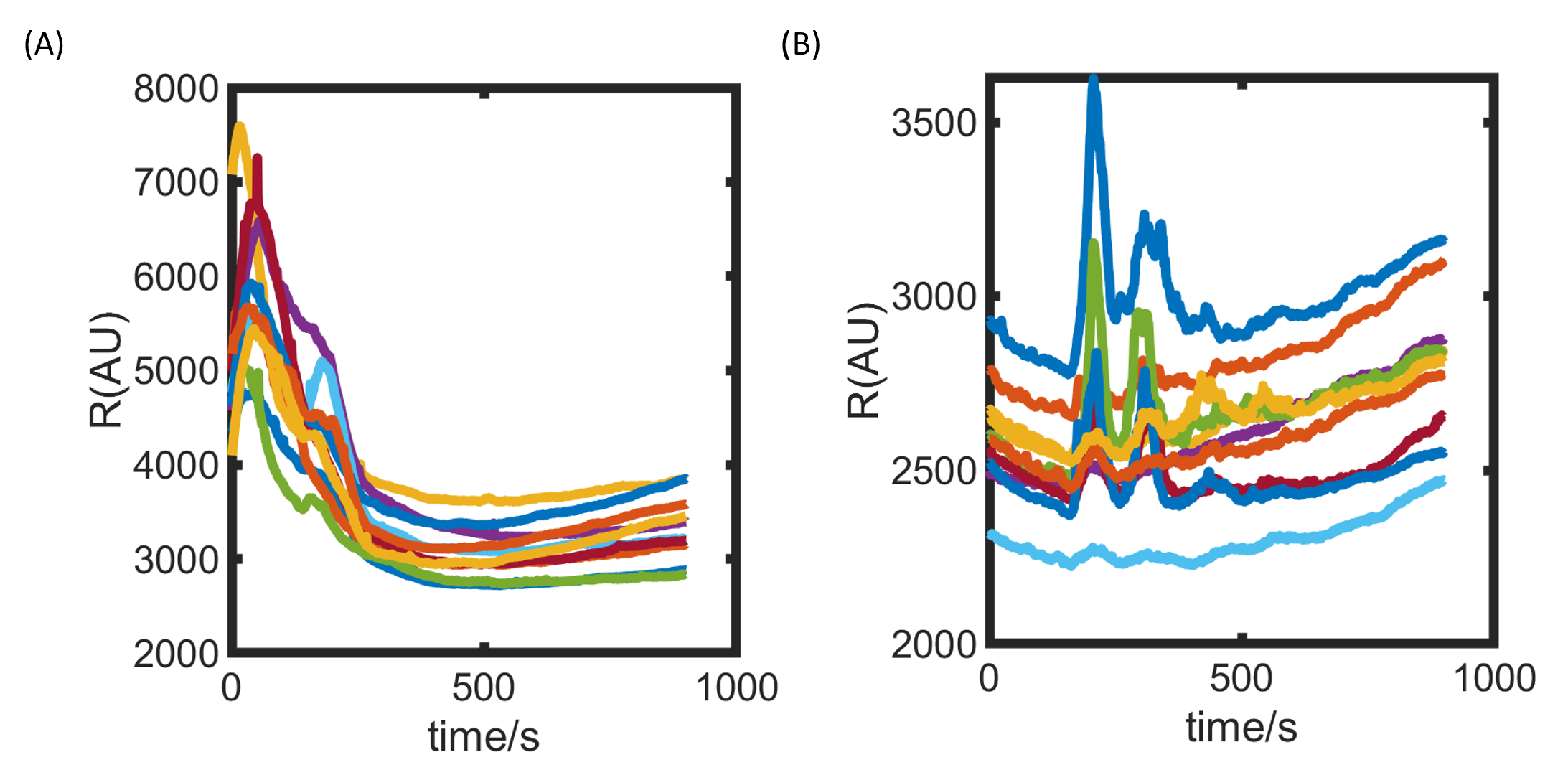}
\caption{Typical calcium responses for thapsigargin treated KTaR cells (A) and KTaR cells in calcium free experiments (B). (A) Fluorescent calcium intensities of 10 randomly selected cells treated by thapsigargin and under 50 $\mu$M ATP at a period of 120 seconds. (D) Fluorescent calcium intensities of 10 randomly selected cells under 50 $\mu$M ATP at a period of 120 seconds. The ATP solution and growth medium are prepared with calcium free reagents. }
\label{SIfig:TP&calcium free}
\end{figure}

Intracellular calcium responses may be the result of calcium influx from extracellular space, or calcium ions released from the endoplasmic reticulum (ER). Here we show that KTaR cell calcium response to repeated ATP stimuli requires both IP3-mediated ER calcium release and extracellular calcium influx.
\bigbreak
\noindent To evaluate the effect of store-operated calcium response, we use thapsigargin to deplete the ER Ca$^{2+}$. Thapsigargin (Thermo Fisher Scientific, Waltham, MA) is first dissolved in DMSO and diluted to 1 $\mu$M in growth medium or ATP solution. When doing thapsigargin treated experiments (50 $\mu$M ATP and 60s+60s switch period), both the ATP solution and growth medium would contain 1 $\mu$M thapsigargin. Thapsigargin is a SERCA pump inhibitor, when cells are treated with thapsigargin, the endoplasmic reticulum can no longer refill calcium ions. The calcium response for thapsigargin treated cells’ experiments show that when SERCA is inhibited, we can only see one or two responses at the beginning of experiments. This indicates that once the calcium ions storage in the endoplasmic reticulum dries out, there are no more response. Combining with our knowledge of ATP stimulation, we believe that the endoplasmic reticulum calcium release dominated by $IP_3$ pathway is responsible for our observed calcium dynamics. The calcium response for the above experiments is plotted in Fig.\ \ref{SIfig:TP&calcium free}.
\bigbreak
\noindent To evaluate the effect of extracellular calcium ion, we conducted 50 $\mu$M ATP, 60s+60s experiments with standard growth medium without calcium ions (calcium free DMEM, high glucose, no glutamine by Thermo Fisher Scientific, Waltham, MA) supplemented with 10\% Fetal Bovine Serum and 2\% 100x penicillin). Cells are cultured with regular growth medium before experiments. The calcium response we observe in calcium free growth medium experiment indicates that extracellular calcium ions are not needed for the first or second cycle calcium response but are necessary for calcium response after 2-3 cycles. This indicates that both ER calcium store and extracellular calcium ions are needed to sustain calcium responses to repeated ATP stimuli.

\subsection*{e.	Manipulating intercellular communications}
Palmitoleic acid (Sigma-Aldrich, MO) is used as a gap junction inhibitor in our experiments. Palmitoleic acid is first dissolved in DMSO and then the solution is diluted to 10 $\mu$M in growth medium. Before the experiment, cells are treated with 10 $\mu$M palmitoleic acid for 8 to 12 hours while seeding into the PDMS device. Then, 10 $\mu$M palmitoleic acid is added to growth medium and ATP solution that used for perfusion. All gap junction inhibited experiments are conducted under 50 $\mu$M ATP, 60s+60s period condition.
\bigbreak
\noindent KCl (potassium chloride, Fisher chemical) is commonly used to trigger action potential on neurons, which lead to electric coupling through gap junctions between neighboring cells. KCl is dissolved in growth medium to make a 30 mM concentration solution. KCl treated cells are compared with 50 $\mu$M ATP experiments under the same period (60s+60s, Fig. 5 of main text).

\section*{S2. Additional information of network structural analysis}
\subsection*{a.	Numerical derivative and stationarity of calcium dynamics}
\begin{figure}[h]
\centering
\includegraphics[width=0.8\textwidth]{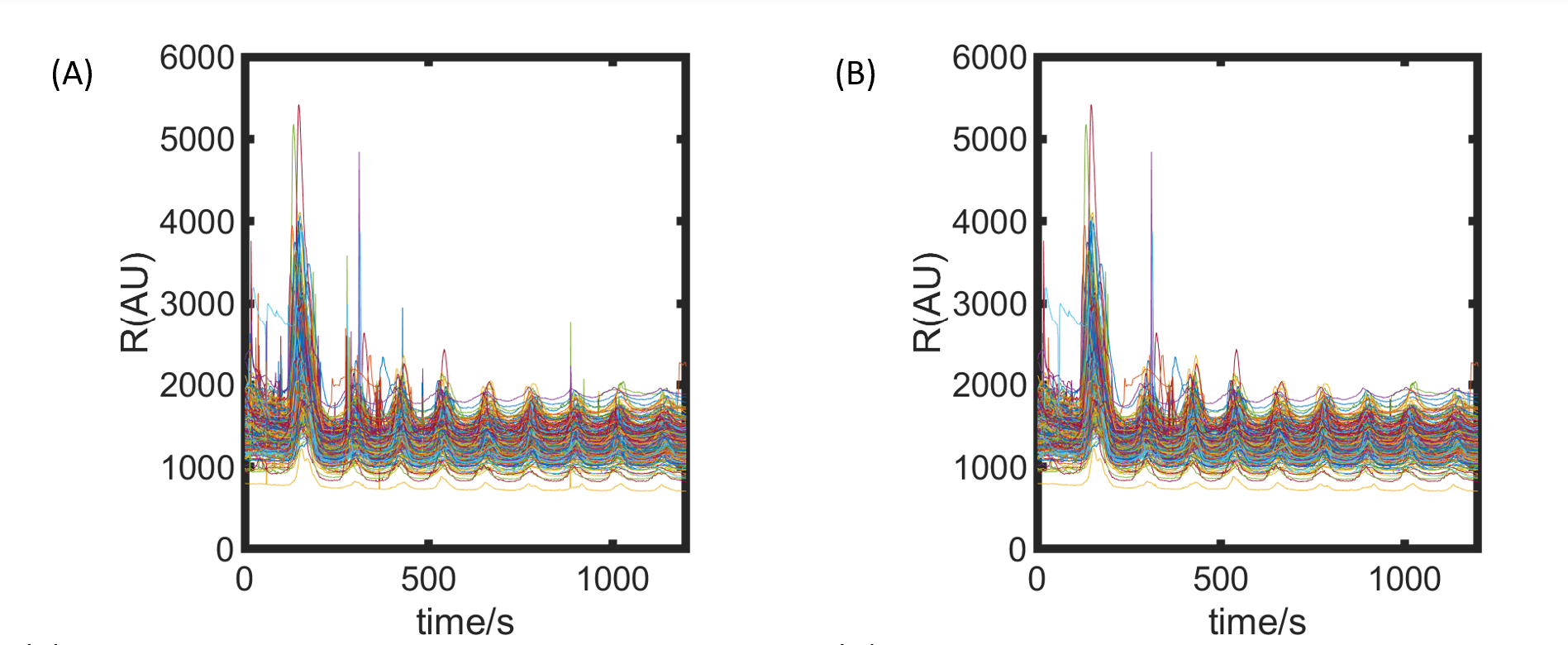}
\includegraphics[width=0.8\textwidth]{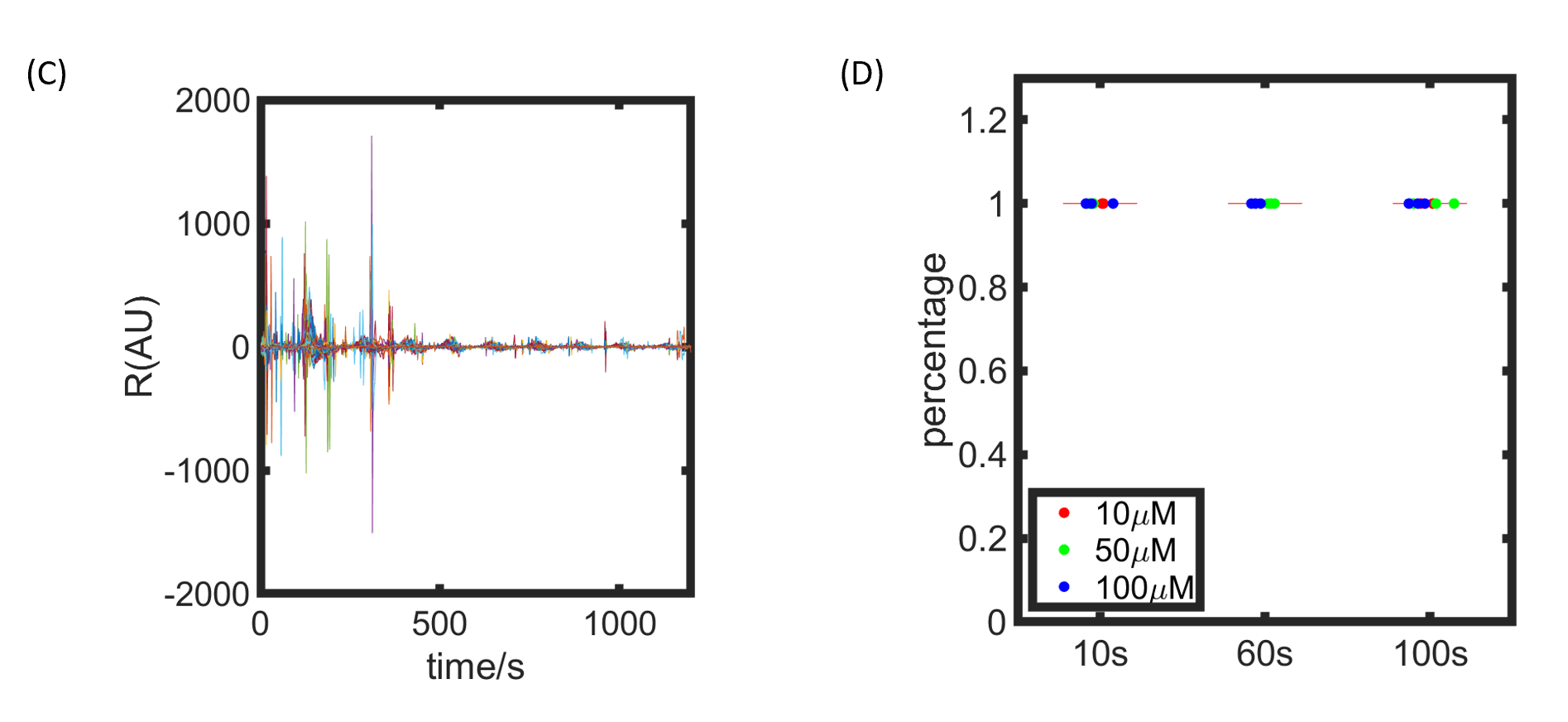}
\caption{Demonstration of fluorescent calcium intensity data after removing outliers and the five-point stencil difference and ADF stationary test result. (A) Typical raw fluorescent calcium intensity data for 50 $\mu$M, 60s+60s experiment. Data in (A), (B) and (C) are all from the same experiment. (B) Fluorescent calcium intensity data after removing outliers. (C) Fluorescent calcium intensity data after removing outliers and after five-point stencil difference. (D) ADF result for fluorescent calcium intensity data from (C). Every cell’s calcium intensity data was tested by adf test, and the result was exhibited as the percentage of cells that passed the test in each experiment. Each dot in the plot represents one experiment.}
\label{SIfigure:stationarytest}
\end{figure}

Individual calcium intensity data obtained from the experiments is first processed in MATLAB before analysis. Meaningless spikes caused by drifting or cell debris will be removed using MATLAB's built-in functions. The outliers were defined as elements more than three local scaled median absolute deviations (MAD) from the local median over a window length of 6 and replaced by a linear interpolation of neighboring, non-outlier values. Then, a five-point stencil difference will be applied to the data to achieve stationary data set. The formula for five-point stencil difference is defined as following:
\begin{equation}
f'(t) = \frac{-f(t+2h)+8f(t+h)-8f(t-h)+f(t-2h)}{12h}
\end{equation}
In the formula, t represents time, h is the time interval which we define as 1 second, f(t) is the intensity data and $f'$(t) is the data after the five-point stencil difference. Data is circularly shifted for calculation at the beginning or the end. The stationarity of the data after five-point stencil difference is checked by using Augmented Dickey-Fuller (adf) test. Fig.\ \ref{SIfigure:stationarytest}A-C shows example of what the data would look like removing outliers and five-point stencil difference. 
\bigbreak
\noindent We used an ADF test to conduct stationarity test for each cell’s fluorescent calcium intensity data before and after the five-point stencil difference in each experiment. We compared the percentage of cells that passed the ADF test and the result indicates that after the five-point stencil, all the cells in all experiments passed the test. Fig.\ \ref{SIfigure:stationarytest}D shows the result of the ADF test for the five-point stencil differenced data.

\subsection*{b.	Granger inference with Granger difference algorithm}
The Granger causality test tests whether one time series is useful in predicting another time series by comparing the prediction ability between two linear auto regression model with the assumptions that (a) cause happens before effect, (b) the cause can help predict the result\cite{Barnett_2015_grangerinneuroscience}. Assuming we have two time series data {x(t)} and {y(t)}, the reduced and full auto regression models testing whether x(t) Granger-cause y(t) at order p is as following:
\begin{equation}
y(t)= \sum_{k=1}^{p} a_1 (t)y(t-k) + \epsilon_1 (t)
\end{equation}
\begin{equation}
y(t)= \sum_{k=1}^{p} a'_1 (t)y(t-k) + \sum_{k=1}^{p} b'_1 (t)x(t-k) + \epsilon_2 (t)
\end{equation}
Inside the formula, $a'_1 (t)$,$b'_1 (t)$  and $a_1 (t)$ are coefficients from linear regression. $\epsilon_1 (t)$ and $\epsilon_2 (t)$ are residuals, p is the order of auto regression model, and the best order will be determined by the Bayesian information criteria (BIC). Using the methods from our reference paper \cite{Belliveau_2014_grangerdifference}, we can calculate the Granger causality metrics using the following formula:
\begin{equation}
GC_{x\rightarrow y} = log(\frac{\sum_{t=1}^{n} \epsilon_1 (t)}{\sum_{t=1}^{n} \epsilon_2 (t)})
\end{equation}
The n in the formula represents the length of the residual. The Granger causality metric will be calculated in both directions (x causes y and y causes x),and the larger one will give us the direction of causality and the difference, which is Granger difference, will be calculated. After that, 5000 surrogate data for the “cause” data are generated by using IAAFT, and the Granger difference between each of the surrogate data and “effect” data are calculated. P value of the test is defined as the percentage of Granger difference of the permuted data that is larger than the Granger difference of the original data. If p <0.05, the causality relation is determined. In our network plotting, we used arrows to indicate the direction and causality between pair of cells.

For each nearest-neighbor pair, we calculate the statistical significance of Granger difference using the time series from a particular cycle. For $T=20$ seconds the corresponding time series is extended to 100 frames to ensure sufficient length for statistical analysis.

\subsection*{c.	Computing Estrada index of the networks}
\begin{figure}[h]
\centering
\includegraphics[width=0.5\textwidth]{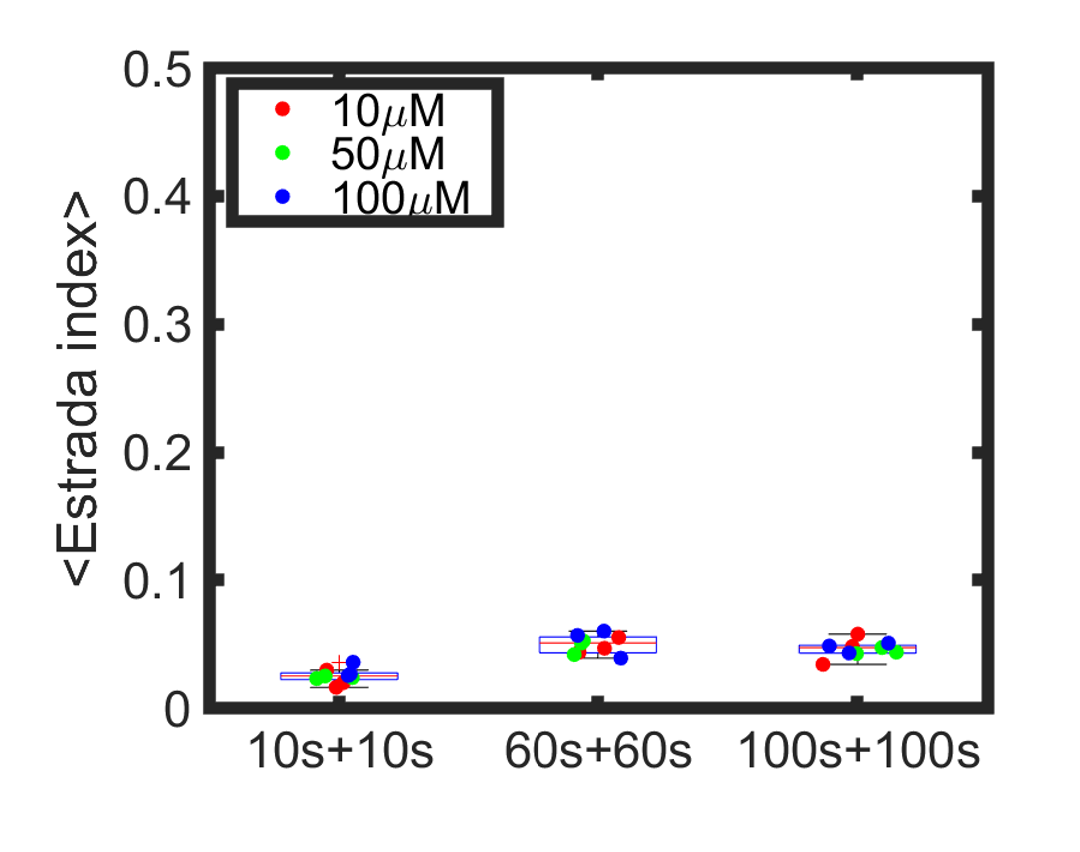}
\caption{The boxplot shows the average Estrada index for each experiment. The Estrada index is calculated by using the normalized formula for each cycle in each experiment. The graph is showing the average value of the normalized Estrada index over all cycles for each experiment. The Labelling on the x axis indicates the stimulation period while the color represents ATP concentration, red green and blue refers to 10 $\mu$M, 50 $\mu$M and 100 $\mu$M. }
\label{SIfig:Estradaindex}
\end{figure}

The Estrada index is calculated based on the following formula \cite{Estrada_2010_estradaindex}. Assume we have node i and j which has degree $k_i$ and $k_j$ (both $k_i$ and $k_j$ are not equal to zero) in the network that has a set of links E and I,J $\in$ E, the Estrada index is defined as following:
\begin{equation}
\rho = \sum_{i,j} (\frac{1}{\sqrt{k_i}}-\frac{1}{\sqrt{k_j}})^2
\end{equation}
And it can be normalized to get a measure $\rho_n$ within interval [0,1]:
\begin{equation}
\rho_n = \frac{\rho}{N-2\sqrt{N-1}}
\end{equation}
Where N represents the total number of nodes. We can calculate the Estrada index for each cycle of experiments and compare the average value of the Estrada index for all of them. In Fig.\ \ref{SIfig:Estradaindex}, we plot the boxplot for visualizing the average value of the Estrada index. From the graph, we can observe that for all experiments, the Estrada index is small (around 0.05) which means that for all different stimulating condition experiments, there is no “star” shape network which means that the network does not have hotspot cells that have overwhelming larger number of connection than other cells

\section*{S3. Additional information of network temporal dynamics}

\subsection*{a.	Additional examples of cycle-to-cycle variation of network structure}
\begin{figure}[h]
\centering
\includegraphics[width=\textwidth]{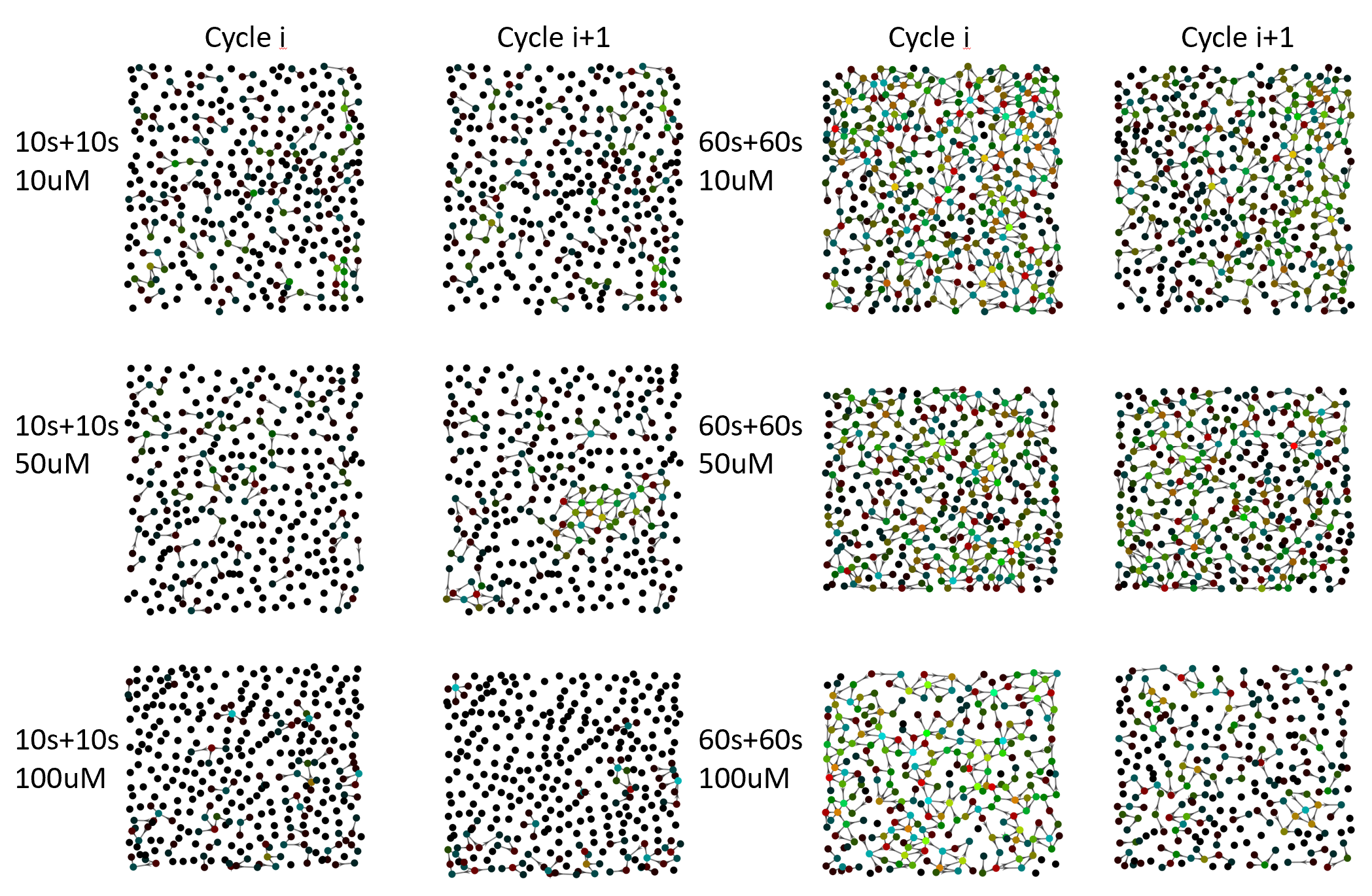}
\includegraphics[width=\textwidth]{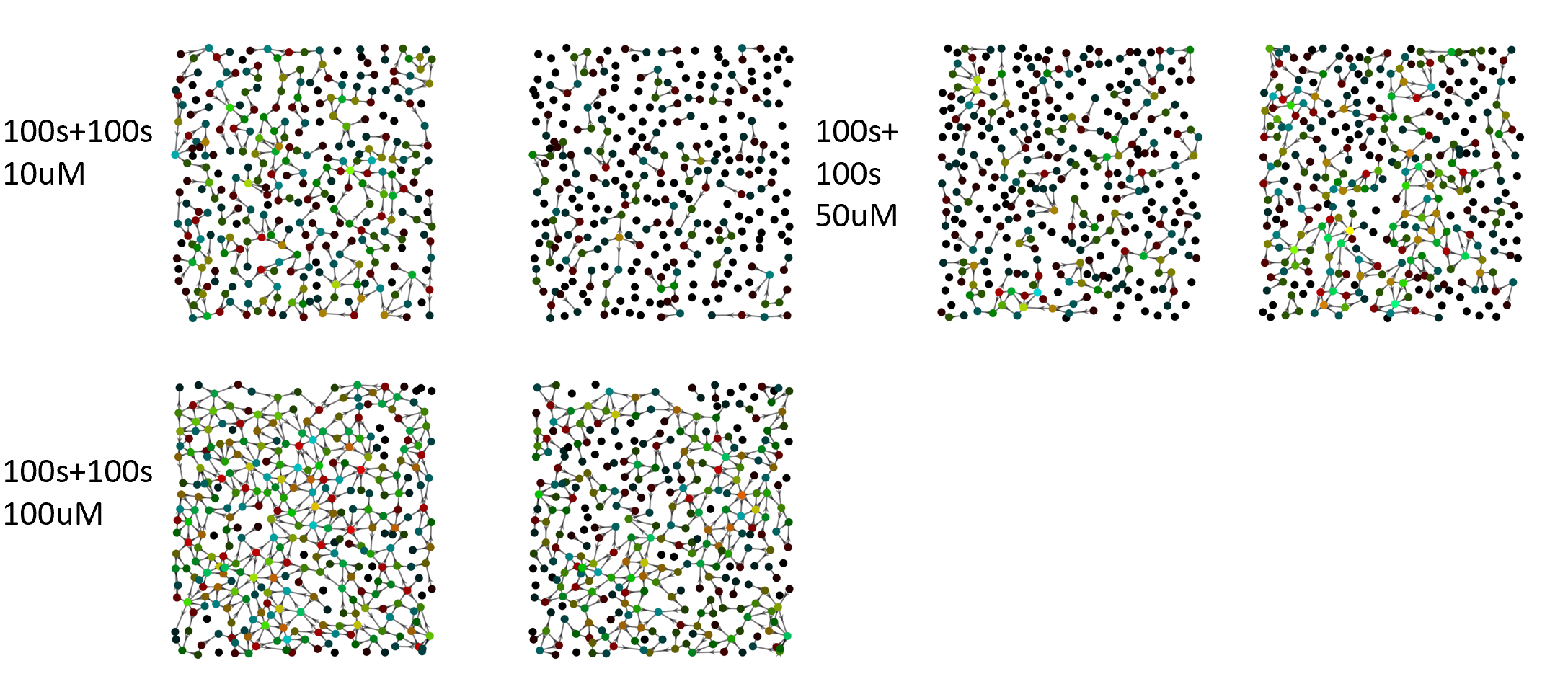}
\caption{Examples of cycle-to-cycle variation of the network for different ATP concentrations and periods. We randomly pick one consecutive cycle from each experiment. The color of the nodes in the graph represents leader scores similar to the main text.}
\label{SIfig:cyclevariation}
\end{figure}

In each cycle, we will construct the network by using the Granger causality test on all the nearest neighbor cell pairs. The length of the intensity data used for the test in each cycle depends on the switching period of experiments. For the 10s+10s experiments, the window size of intensity data is 100s (0 to 100s, 20 to 120s…), while for 60s+60s and 100s+100s experiments, the window size equals to the switch period, which is 120s and 200s separately. In Fig.\ \ref{SIfig:cyclevariation}, we can observe more network’s cycle-to-cycle variation.

\section*{b.	Network transition rates do not depend on ATP concentration}
\begin{figure}[h]
\centering
\includegraphics[width=\textwidth]{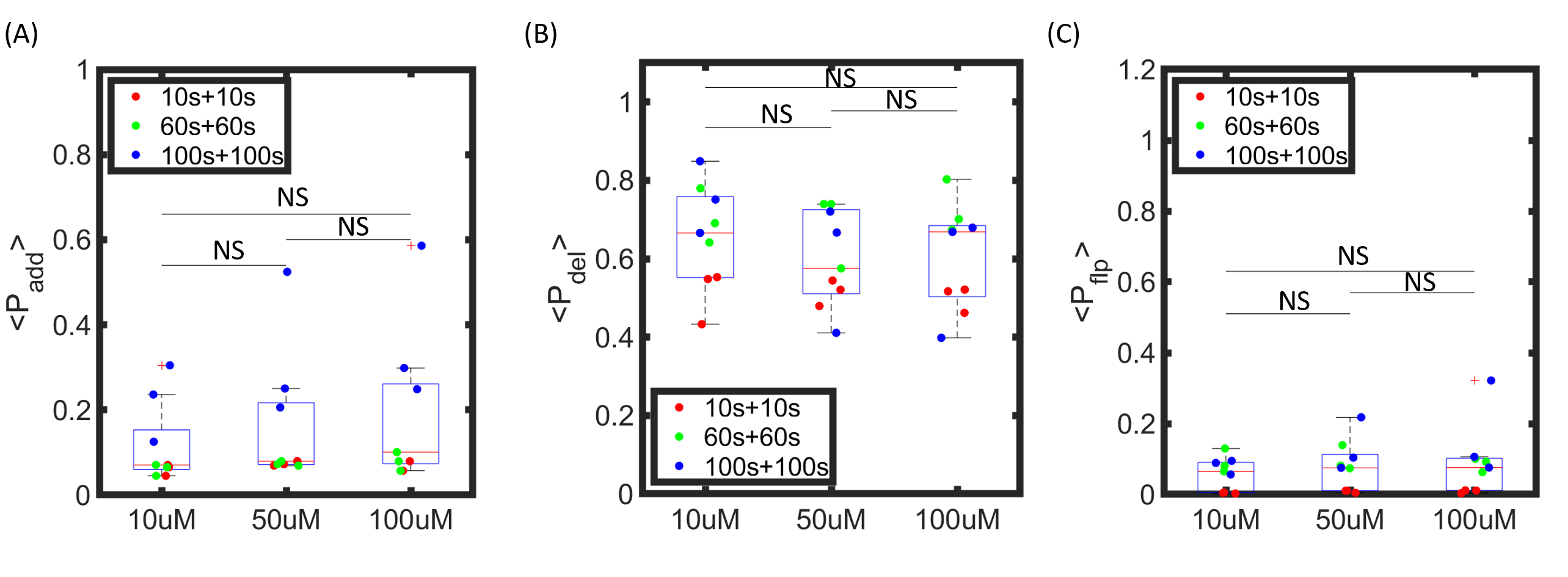}
\caption{The boxplots show the dependence of network kinetics on ATP concentration. (A) The dependence of $P_{add}$ (averaged over cycles per each experiment, as explained in the main text) on ATP concentration. (B) The dependence of average $P_{del}$ on ATP concentration. (C) The dependence of average $P_{flp}$ on ATP concentration. In each boxplot, the color represents periods (red, green, blue for 10s+10s, 60s+60s and 100s+100s respectively). Statistical comparisons are done with Tukey’s honest significant test after one way ANOVA. **: p < 0.01, ***: p < 0.001, n.s. : not significant.}
\label{SIfig:concentrationPaddPdelPflp}
\end{figure}

As shown in Fig.\ \ref{SIfig:concentrationPaddPdelPflp}, we find the network kinetics, as characterized by the transition rates $P_{add}$, $P_{del}$, and $P_{flp}$, do not depend on ATP concentration. 

\section*{c.	Network transition rates do not depend on leader/follower scores}
We perform a computational simulation to test the null hypothesis that Network transition rates $\{P_{add}, P_{del}, P_{flp}\}$ do not depend on leader/follower scores. In particular, we take the following steps:

\begin{enumerate}
  \item Choose one experiment as a template, network of the first cycle is directly copied from the experiment
  \item Each edge is evolved independently without regarding local leader/follower scores. The average rates (reported in the main text) are used to evolve the network for the same number of cycles as experiments.
  \item For the edges existing in previous cycle, the possibility for removing the edge is equal to $P_{del}$ from experiments. For edges remain in the next cycle, the possibility for flipping direction equals to $P_{flp}$ calculated from experiments.
  \item For empty sites in the previous cycle, the possibility for constructing a new edge is equal to $P_{add}$ from the experiments. If a new edge is constructed, the probability for generating the edge in an outward or an inward direction will be the same (0.5 and 0.5 respectively).
\end{enumerate}

\noindent Based on the rules, we choose one experiment for each period for simulation. For each iteration, the number of networks that we generated is equal to the number of cycles in real experiments. For each experiment, we perform 10,000 iterations of numerical simulations.

\noindent Using numerical simulations, we compute the distribution of transition rates conditioned with local leader/follower scores. An example is given in Fig.\ \ref{SIfig:paddpdel}A where the numerical simulations are derived from experiments with $T=20$ sec and ATP concentration of 50 $\mu$M. Red lines indicate 95\% confidence interval and the black dashed line indicates experimental result. This data shows that the experimentally measured $P_{add}$ is consistent (within the confidence interval) with null hypothesis. Fig.\ \ref{SIfig:paddpdel}B-C show the analysis 3 experiments at varying periods. It is clear that the experimental results are consistent with the null hypothesis: network transition rates do not depend on leader/follower scores. 

\begin{figure}[h]
\centering
\includegraphics[width=0.5\textwidth]{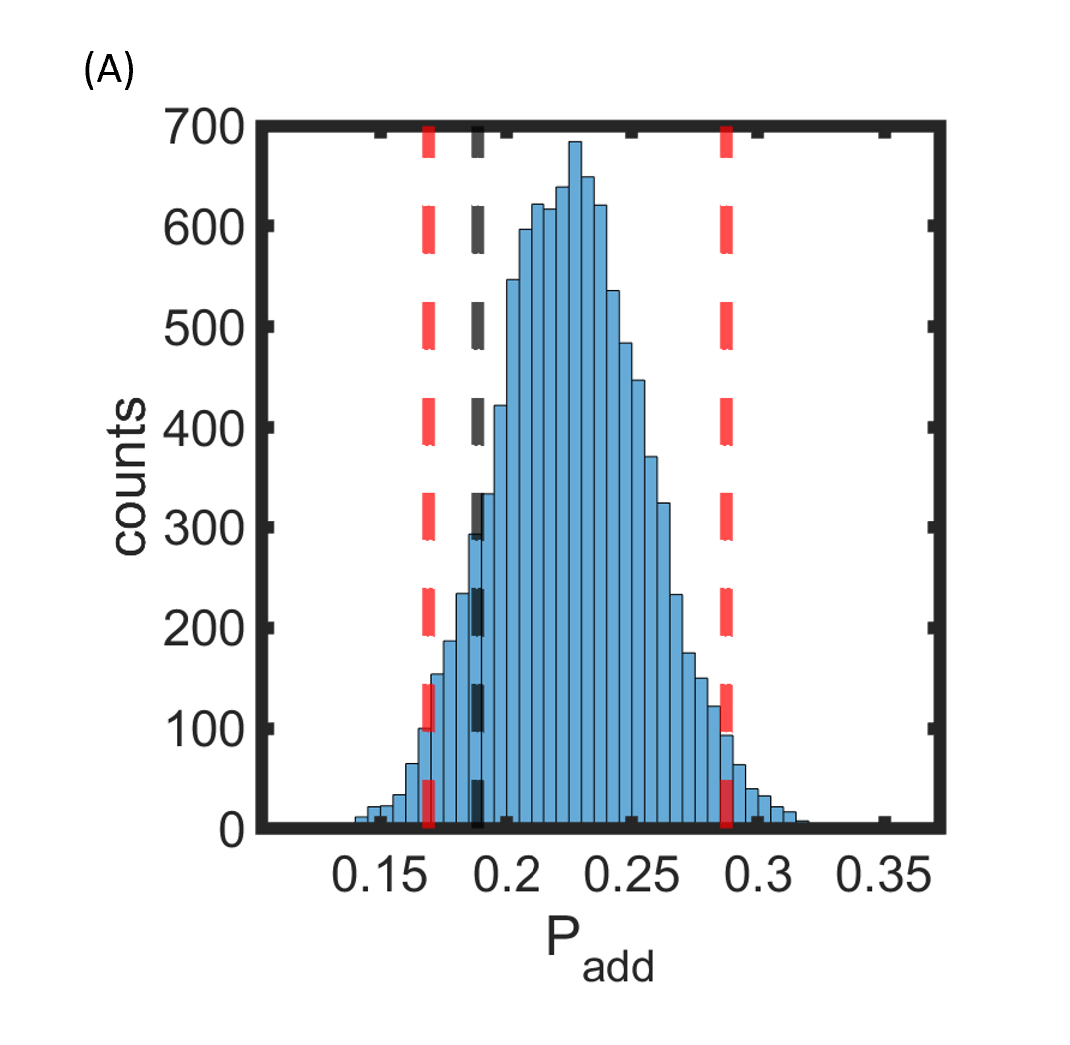}
\end{figure}

\begin{figure}[h]
\centering
\includegraphics[width=\textwidth]{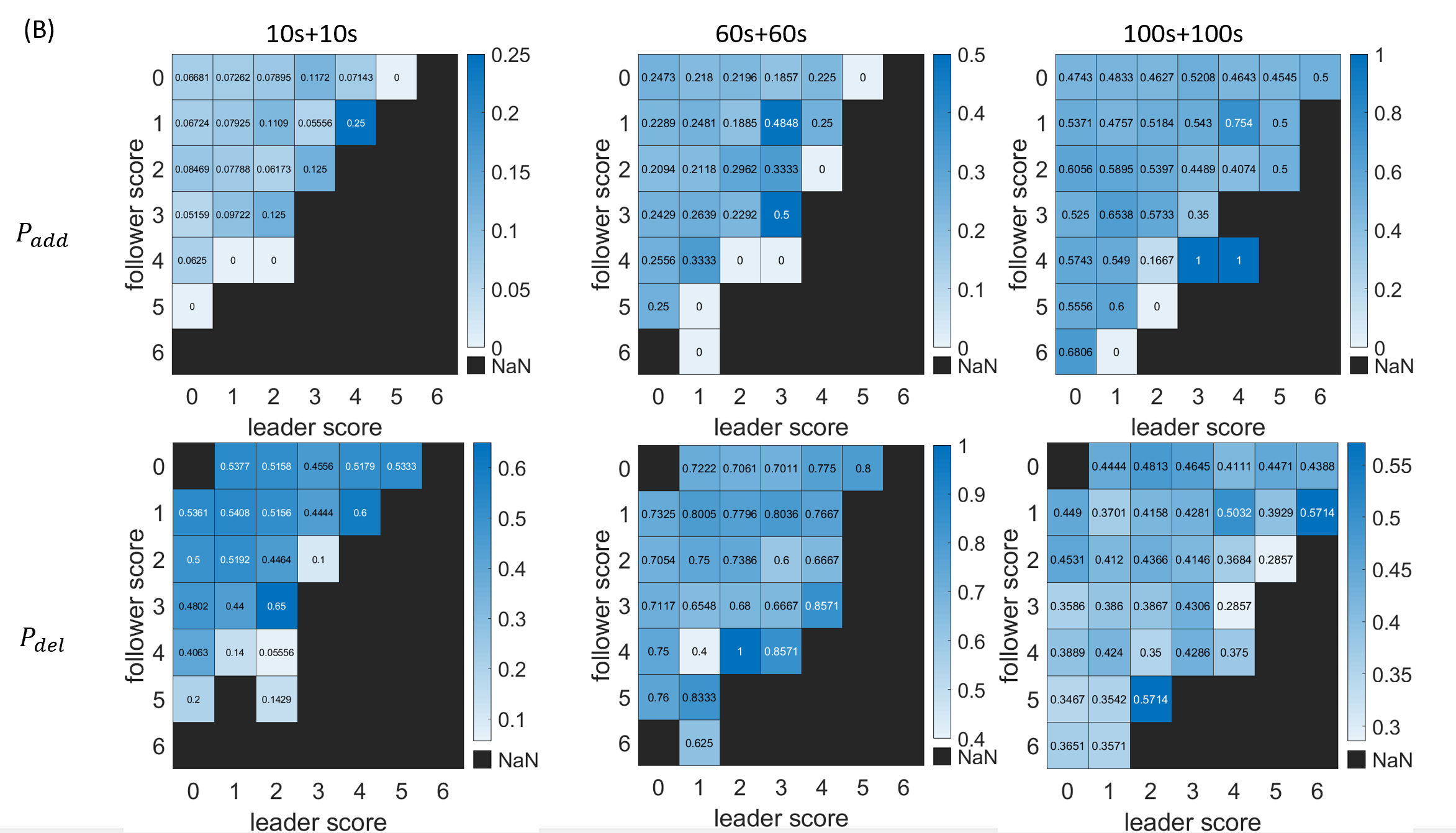}
\includegraphics[width=\textwidth]{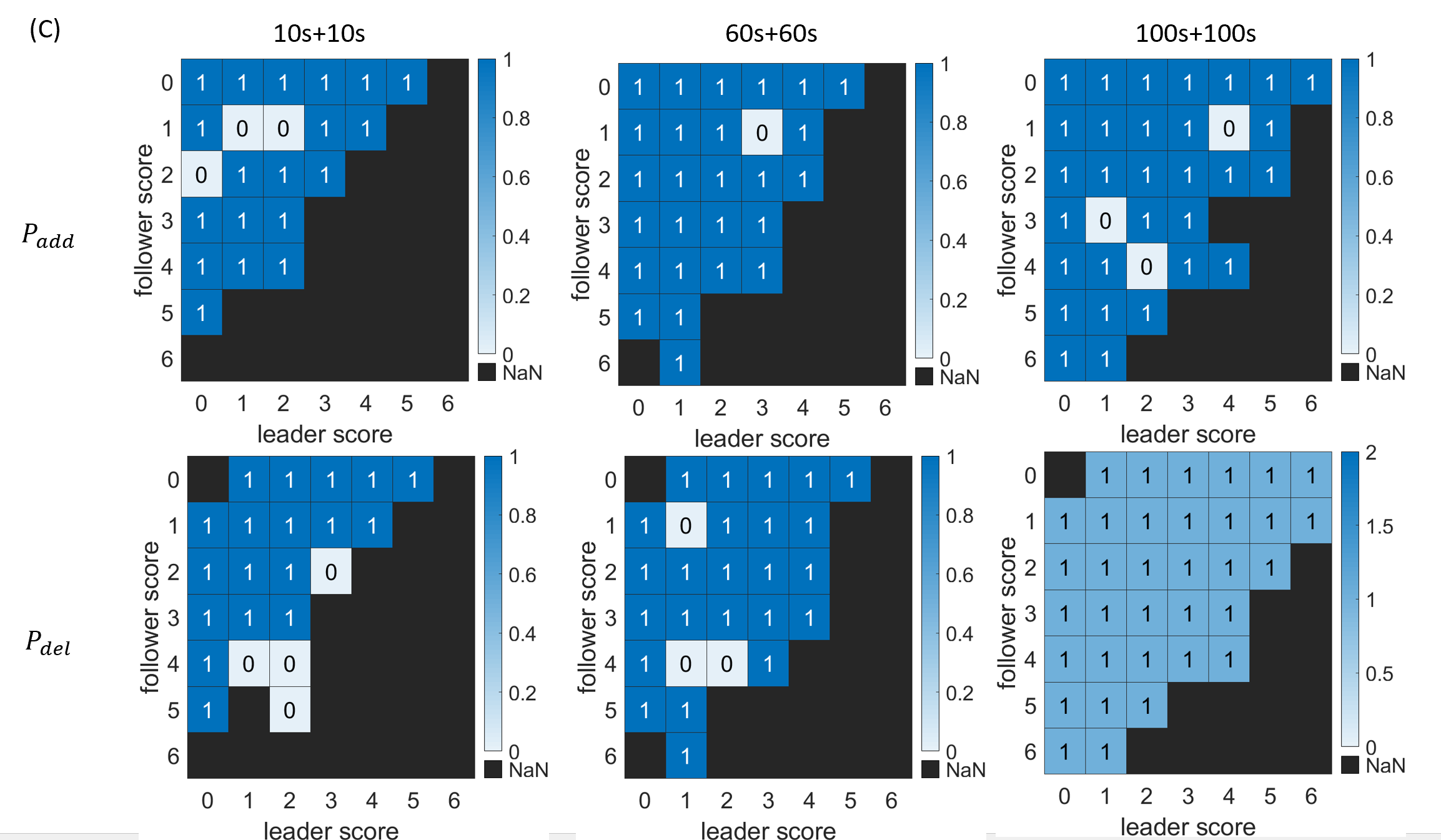}
\caption{Testing the hypothesis that $P_{add}$ and $P_{del}$ do not depend on individual cells classified by leader/follower scores. We choose ATP concentration of 50 $\mu$M but examine all periods. (A) From the 10000 iterations of simulated network based on a $T=$ 20 sec, $[ATP]=$ 50 $\mu$M experiment, we calculate the distribution of $P_{add}$ for cells having leader score = 2, follower score = 3 (blue bar). The red dotted line represents the upper and lower boundary of the 95\% confidence interval while the black dotted line is the data from experimental measurement. (B) Experimental measurement of average $P_{add}$ and $P_{del}$ for individual cells classified by leader/follower scores. Each heatmap represents one experiment. From left to right, the period for the experiment would be 10s+10s, 60s+60s and 100s+100s. (C) Result of whether the experimental measurement of average $P_{add}$ and $P_{del}$ for individual cell are inside the 95\% confidence interval. Heatmaps in (C) resembles the heatmaps in (B) but only have number 1 or 0. If the number in the block is 1, it means that the $P_{add}$ ($P_{del}$) at the corresponding block is within the confidence interval while 0 means the opposite.}
\label{SIfig:paddpdel}
\end{figure}

\section*{d.Network reaches dynamic equilibrium and follows detailed balance}
\begin{figure}[h]
\centering
\includegraphics[width=0.7\textwidth]{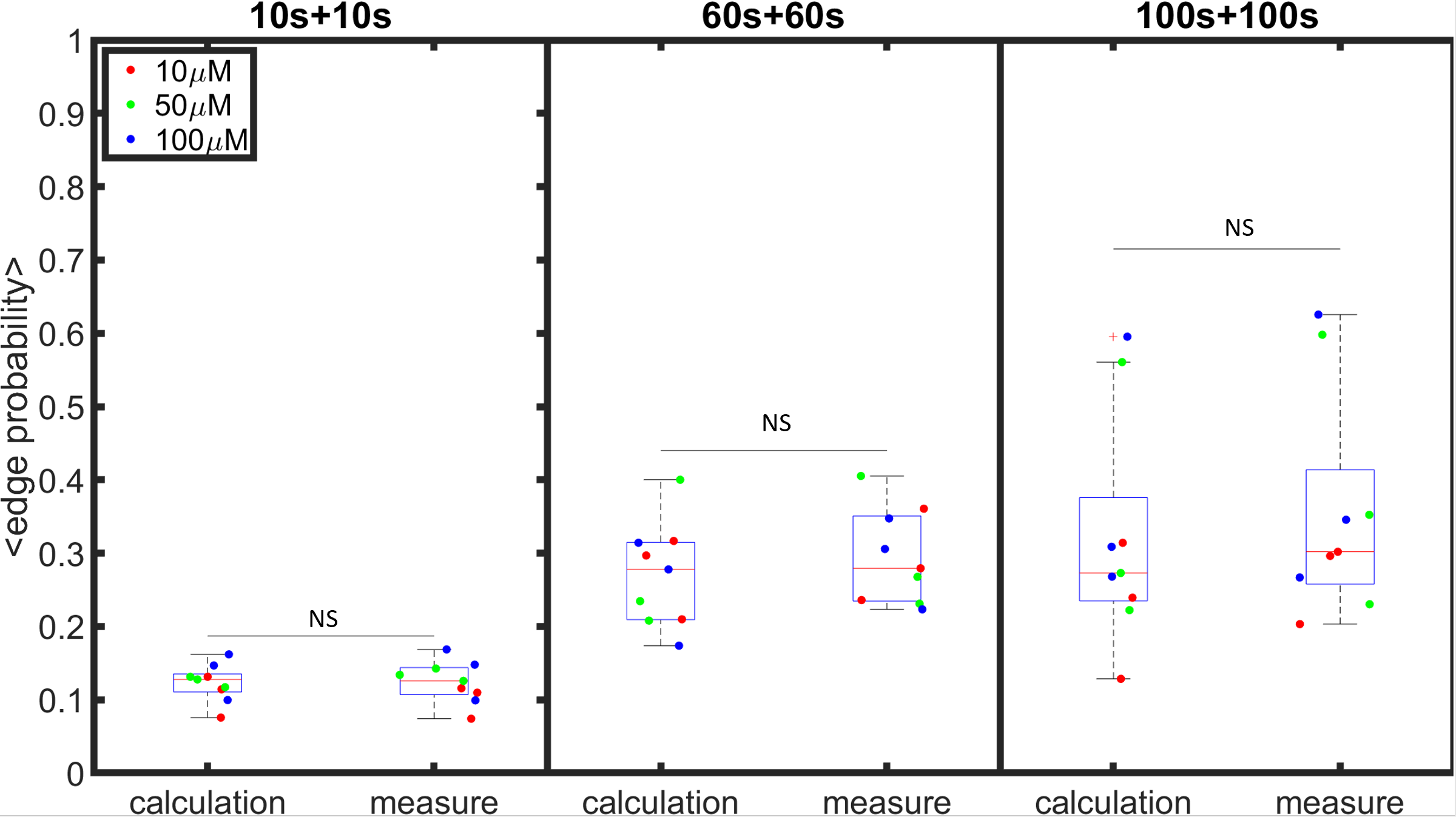}
\caption{Comparison between calculated edge probability and measured edge probability. Each dot represents one experiment. The plot is divided into three parts, each part represents one switch period while the color represents ATP concentration (red, green, blue for 10 $\mu$M, 50 $\mu$M, 100 $\mu$M). Statistical comparisons are done with Tukey’s honest significant test after one way ANOVA. **: p < 0.01, ***: p < 0.001, n.s. : not significant. }
\label{SIfig:edgeprob_calculation}
\end{figure}

Following the definition of $P_{add}$ and $P_{del}$ from the main text, if the network structure is stationary over consecutive cycles, we will be able to derive the following equation for each cycle:
\begin{equation}
    P_{add} * (N - N_{edges}) = P_{del} * N_{edges}
\end{equation}
In the formula, N represents the total number of nearest neighbor pairs, $N_{edges}$ represents the number of edges for that cycle. Then, dividing both sides of the equation by N, we will have:
\begin{equation}
    P_{add} * (1 - P_{edges}) = P_{del} * P_{edges}
\end{equation}
Inside the formula, $P_{edges}$ represents the edge probability. We can then calculate the edge probability using the following equation:
\begin{equation}
    P_{edges} = \frac{P_{add}}{P_{del}+P_{add}}
\end{equation}
Using the average $P_{add}$ and $P_{del}$ for each experiment, we can calculate the average edge probability using the formula and compare it with experimental data. Fig.\ \ref{SIfig:edgeprob_calculation} shows the result of comparison between simulated data and original data. The multi-comparison test indicates that there is no difference between calculation and experimentally measured edge probability, indicating that the network is indeed in dynamic equilibrium.

\begin{figure}[h]
\centering
\includegraphics[width=0.5\textwidth]{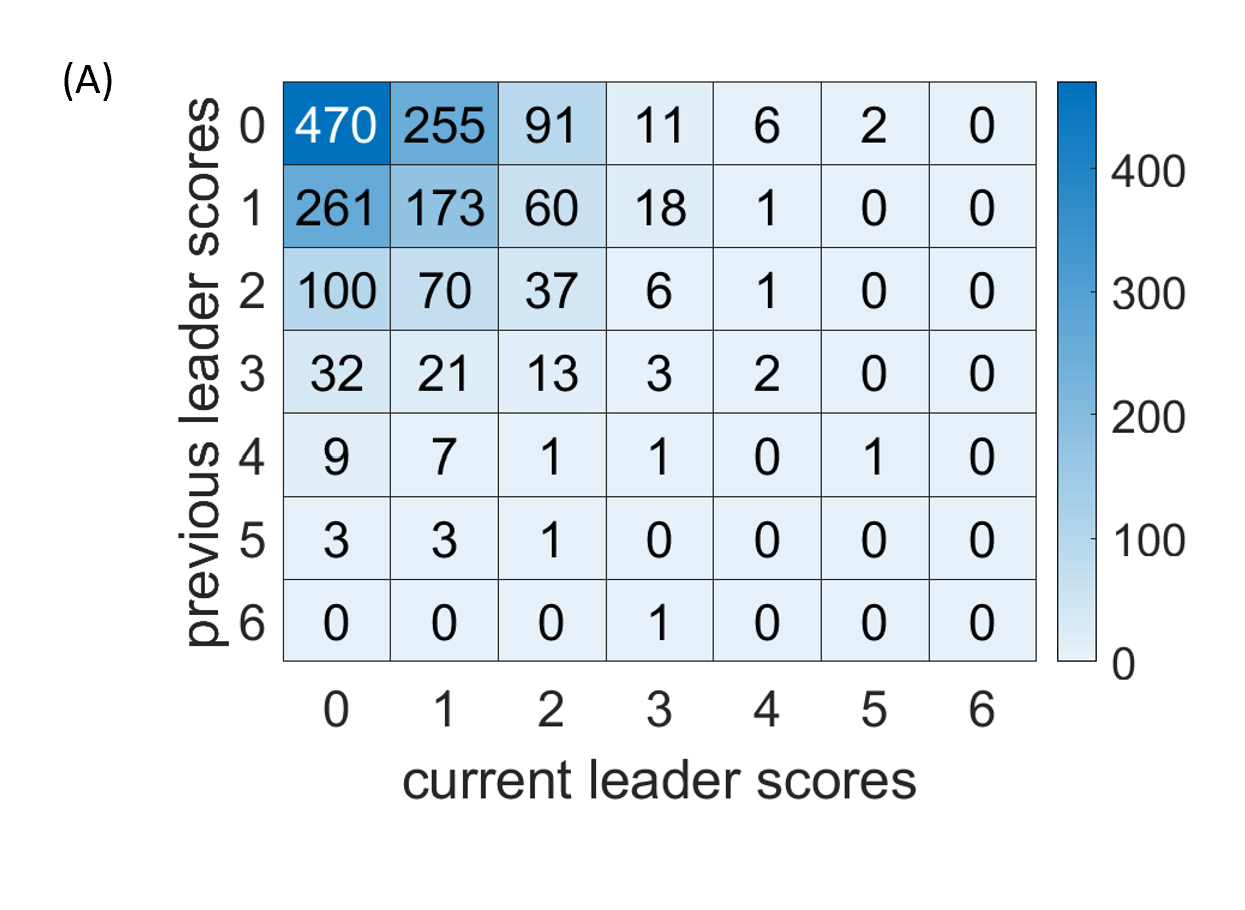}
\includegraphics[width=0.9\textwidth]{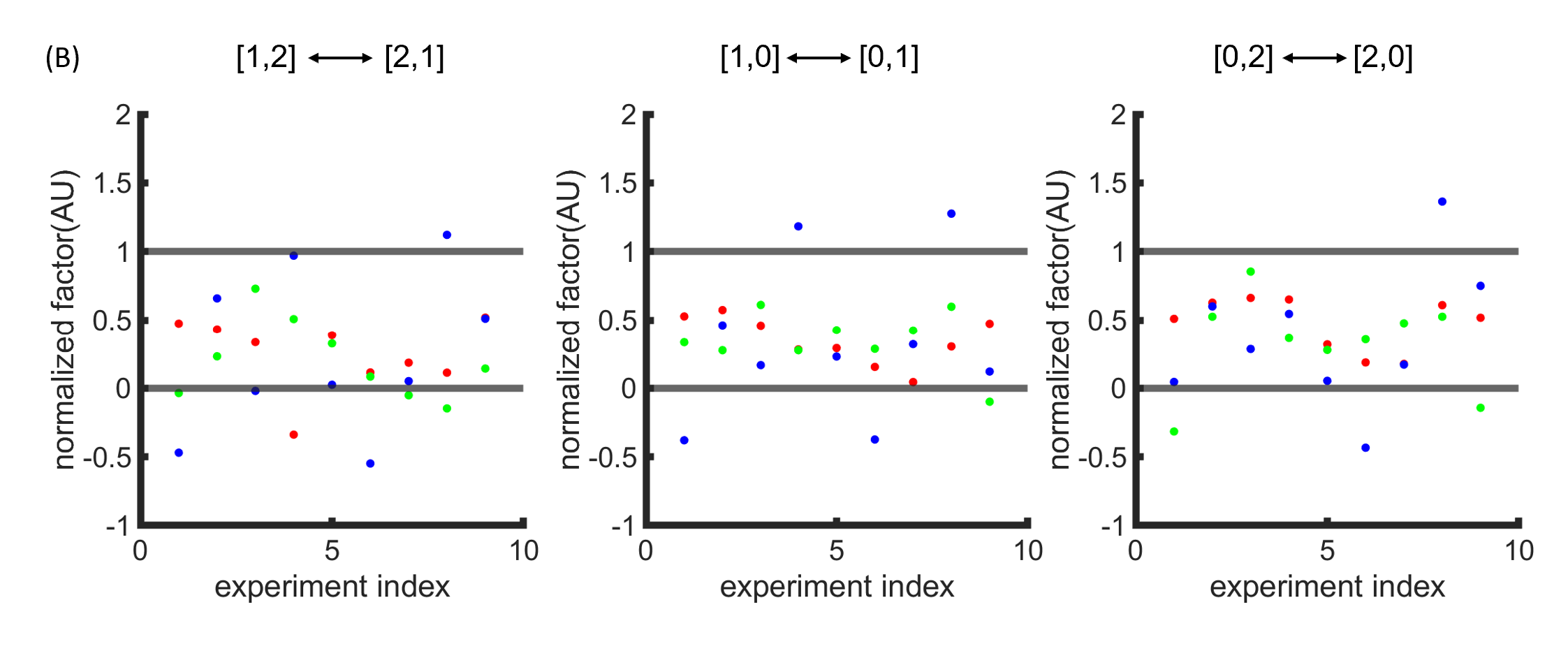}
\caption{Demonstration of detailed balance in the cellular state space defined by the leader score. (A) An experimentally measured event matrix showing the total number of transitions between previous and current leader scores. The corresponding experiment has KTaR cells experiencing a period of 60s+60s and 50 $\mu$M ATP stimuli. (B) The normalized off-diagonal differences compared with 95\% confidence intervals of detailed balance. See text for more details. Each dot represents one experiments and red, green and blue corresponds to periods of 10s+10s, 60s+60s, 100s+100s. Black dashed lines indicating the boundary of 95\% confidence interval. }
\label{SIfig:detailedbalance}
\end{figure}

\bigbreak
\noindent Considering leader/follower scores as labels of cell states, we next investigate if the cellular state transitions in consecutive cycles follow detailed balance. If they do, it is expected entropy will be produced in the cell state space. To this end, we compute an event matrix for each experiment. An event matrix is generated by counting the number of transitions between previous and current leader score from a particular experiment (the entries are also known as fluxes, Fig.\ \ref{SIfig:detailedbalance}A). Next we will show that the event matrix is statistically consistent with a symmetric matrix, indicating detailed balance (no net flux) in the leader score space. Same can be shown for the event matrix of follower scores.

\noindent To avoid bias due to rare transition which leads to small number of events, we will only focus on the top left 3x3 block whose previous and current leader score range from 0 to 2. For all experiments, the top left area will account for 80-90\% of all events. By doing so, we will only have 3 off-diagonal term pairs to check: using the format of [ previous leader score, current leader score] we have [1,0] $\longleftrightarrow$ [0,1]; [2,0] $\longleftrightarrow$ [0,2]; [1,2] $\longleftrightarrow$ [2,1]. 

Although a experimentally measured event matrix (Fig.\ \ref{SIfig:detailedbalance}A) is slightly non-symmetric, it could be due to finite sample size effect. To test the null hypothesis that the cells follow detailed balance in leader score space, we simulate transition events based on symmetrized experimental event matrix. In particular, if $T_{exp}$ is a experimentally measured event matrix (such as Fig.\ \ref{SIfig:detailedbalance}A), then the simulation follows probabilities calculated from $\frac{1}{2}(T_{exp}+T_{exp}^t)$. The total number of events in each iteration of simulation is fixed to be the same as the corresponding experiment. For each experiment, 1000 iterations of simulation are performed and the distribution of off-diagonal terms' differences are calculated. 

We next examine if the experimentally measured off-diagonal difference is within the 95\% confidence interval of our null hypothesis. To better compare across different experiments, we linearly normalize the upper and lower bounds of the confidence interval for each experiment to 1 and 0, and normalize the experimental values in the same way: $N_{exp} \rightarrow$ normalized factor $= \frac{N_{exp}-N_{lower}}{N_{upper}-N_{lower}}$.

As shown in  \ref{SIfig:detailedbalance}B,  nearly all of the off-diagonal term difference from our experimental data are within the 95\% confidence interval. These result suggest that there exists detailed balance in the cellular state space (as defined by leader score).

\section*{e.	Continuous wavelet transformation analysis}
In our analysis, we conduct a continuous wavelet transformation analysis using MATLAB. The continuous wavelet transformation is a time-frequency analysis. The formula we use is for continuous 1D wavelet transformation as following:
\begin{equation}
    F(\tau,s) = \frac{1}{\sqrt{s}} \int_{-\infty}^{+\infty} f(t)\phi(\frac{t-\tau}{s}) \,dt\
\end{equation}
Inside the formula, $\phi$ is the mother wavelet function, a is scaling which is related with frequency, it dominates the stretch or compress of the wavelet function, $\tau$ is the time delay, and F will be the continuous wavelet transformation coefficient. 
\bigbreak
\noindent In our analysis, we use the MATLAB built-in function “cwt” for wavelet transformation analysis, Morse wavelet is chosen to be our mother wavelet function with its symmetry parameter ($\gamma$) equal to 3 and its time-bandwidth product equal to 60. 
\bigbreak
\noindent Based on the continuous wavelet transformation(cwt) coefficient, we define normalized wavelet score (NWS) to check if cells are responding at the input frequency. In our experiments, the frequency of a input stimulation is 1/120 Hz = 0.0083 Hz. Thus, the reference frequency ($f_{ref}$) is defined as a range between 0.008 to 0.0089 Hz. If continuous wavelet transformation analysis gives us multiple frequencies within that range, we will choose the one with highest cwt coefficient.
\bigbreak
\noindent The NWS is calculated by using the cwt coefficient at a reference frequency divided by the maximum cwt coefficient with time t, which is shown by the following formula:
\begin{equation}
    NWS(t) = \frac{F(f_{ref},t)}{max(F(f,t))}
\end{equation}
In that formula, F represents the cwt coefficient.  With this definition, if the calcium dynamics of a cell perfectly follows the driving frequency ($f_{ref}$), its NWS equals 1 at all times (except for boundary effects that affect the beginning and end of the time series). Otherwise, the NWS will fluctuate between 0 and 1 when irregular response occurs.

\section*{S4. Theoretical Modeling}

\subsection*{a. Calcium Dynamics Model}

Inspired by the FitzHugh-Nagumo model and wanting an inherently noisy system we 
have come up with a set of reactions to model the calcium dynamics in excitable cells. 

\begin{equation}
    \emptyset \mathrel{\substack{k_1\\\rightleftarrows\\ k_2}} X, \quad
    2X \mathrel{\substack{k_3\\\rightleftarrows\\ k_4}} 3X, \quad 
    X \stackrel{k_6}{\rightarrow} Y + X, \quad 
    Y + X \stackrel{k_5}{\rightarrow} Y, \quad 
    Y \stackrel{k_7}{\rightarrow} \emptyset
    \label{reactions}
\end{equation}
The following two rate equations can be derived from the reactions.
\begin{equation}
    \frac{dx}{dt} = \dot{x}=k_1 - k_2x + k_3x^2 - k_4x^3 - k_5xy
    \label{x_dot}
\end{equation}
\begin{equation}
    \frac{dy}{dt} = \dot{y} = k_6x - k_7y
    \label{y_dot}
\end{equation}
In its current form it is not exactly clear how the parameters affect the system. To get a better understanding,
the system is transformed into something more familiar. The steps to do so are: 

1) Define new variables

2) Eliminate the quadratic term at the fixed point

3) Define new parameters based on the Landau theory of the Ising model
\\

We begin by defining two new variables:
\begin{equation}
    m \equiv \frac{x - x_c}{x_c}
    \label{m}
\end{equation}
\begin{equation}
    n \equiv \frac{y - \frac{k_6}{k_7}x}{x_c}\
    \label{n}
\end{equation}
Substituting Eq.\ \ref{m} and Eq.\ \ref{n} into Eq.\ \ref{x_dot} and Eq.\ \ref{y_dot} and solving for $\dot m$ and 
$\dot n$ we get a new set of equations.
\begin{equation}
    \frac{dm}{dt} = \dot m = \frac{k_1}{x_c} - k_2(m+1) + (\frac{k_3 k_7 - k_5 k_6}{k_7})x_c(m+1)^2 - 
    k_4 x_c^2(m+1)^3 - k_5 x_c(m+1)n
    \label{m_dot}
\end{equation}
\begin{equation}
    \frac{dn}{dt} = \dot n = - k_7 n - \frac{k_6}{k_7}\dot m
    \label{n_dot}
\end{equation}
The fixed points of a system exist where the nullclines cross. The nullclines of Equations \ref{m_dot} and 
\ref{n_dot} are respectively,

\begin{equation}
    n = \frac{\frac{k_1}{x_c} - k_2 (m+1) + (\frac{k_3 k_7 - k_5 k_6}{k_7})x_c (m+1)^2 - 
    k_4 x_c^2(m+1)^3}{k_5 x_c (m+1)}
    \label{m_null}
\end{equation}
\begin{equation}
    n = \frac{-\frac{k_6}{k_7}(\frac{k_1}{x_c} - k_2 (m+1) + (\frac{k_3 k_7 - k_5 k_6}{k_7})x_c (m+1)^2 - 
    k_4 x_c^2(m+1)^3)}{k_7 - \frac{k_5 k_6}{k_7} x_c (m+1)}.
    \label{n_null}
\end{equation}
These nullclines cross when
\begin{equation}
    \frac{k_1}{x_c} - k_2 (m+1) + (\frac{k_3 k_7 - k_5 k_6}{k_7})x_c(m+1)^2 - k_4x_c^2(m+1)^3 = 0.
    \label{fixedpt1}
\end{equation}
\newline

We now eliminate the quadratic term by defining $x_c$ such that
\begin{equation}  
    x_c = \frac{k_3 k_7 - k_5 k_6}{3 k_4 k_7}
    \label{x_c}
\end{equation}
Substituting Eq.\ \ref{x_c} into Eq.\ \ref{fixedpt1} and dividing by $3k_4x_c^2$ gives
\begin{equation}
    (\frac{k_1}{3 k_4 x_c^3} - \frac{k_2}{3 k_4 x_c^2} + \frac{2}{3}) - (\frac{k_2}{3 k_4 x_c^2} - 1)m - 
    \frac{1}{3}m^3 = 0
    \label{fixedpt2}
\end{equation}
Defining 
\begin{equation}
    h \equiv \frac{k_1}{3 k_4 x_c^3} - \frac{k_2}{3 k_4 x_c^2} + \frac{2}{3}
    \label{h}
\end{equation}
\begin{equation}
    \theta \equiv \frac{k_2}{3 k_4 x_c^2} - 1
    \label{theta}
\end{equation}
Equation \ref{fixedpt2} becomes
\begin{equation}
    h - \theta m - \frac{1}{3}m^3 = 0
    \label{fixedpt3}
\end{equation}
Equation \ref{fixedpt3} is now in the form of the minima of the Landau free energy of the Ising model \cite{goldenfeld}. $h$ can be regarded as an external field and $\theta$ as a reduced temperature.
This form is more interpretable for what happens to the fixed point as the parameters are changed. 
\newline

Putting equations \ref{m_dot} and \ref{n_dot} in terms of $h$ and $\theta$ the system becomes
\begin{equation}
    \frac{dm}{d\tau} = \dot m = h - \theta m - \frac{1}{3} m^3 - \phi (m+1)n
    \label{m_dot2}
\end{equation} 
\begin{equation}
    \frac{dn}{d\tau} = \dot n = -k_7 (\theta + 1) n - \frac{k_6}{k_7}(\theta + 1)\dot m
    \label{n_dot2}
\end{equation}
where 
$$\tau = 3 k_4 x_c^2$$
$$\phi = k_5 x_c$$
The nullclines for this system of equations is plotted below in Figure \ref{nullclines} with varying parameters.
\\
\begin{figure}[h][ht]
    \centering
    \includegraphics[scale = .75]{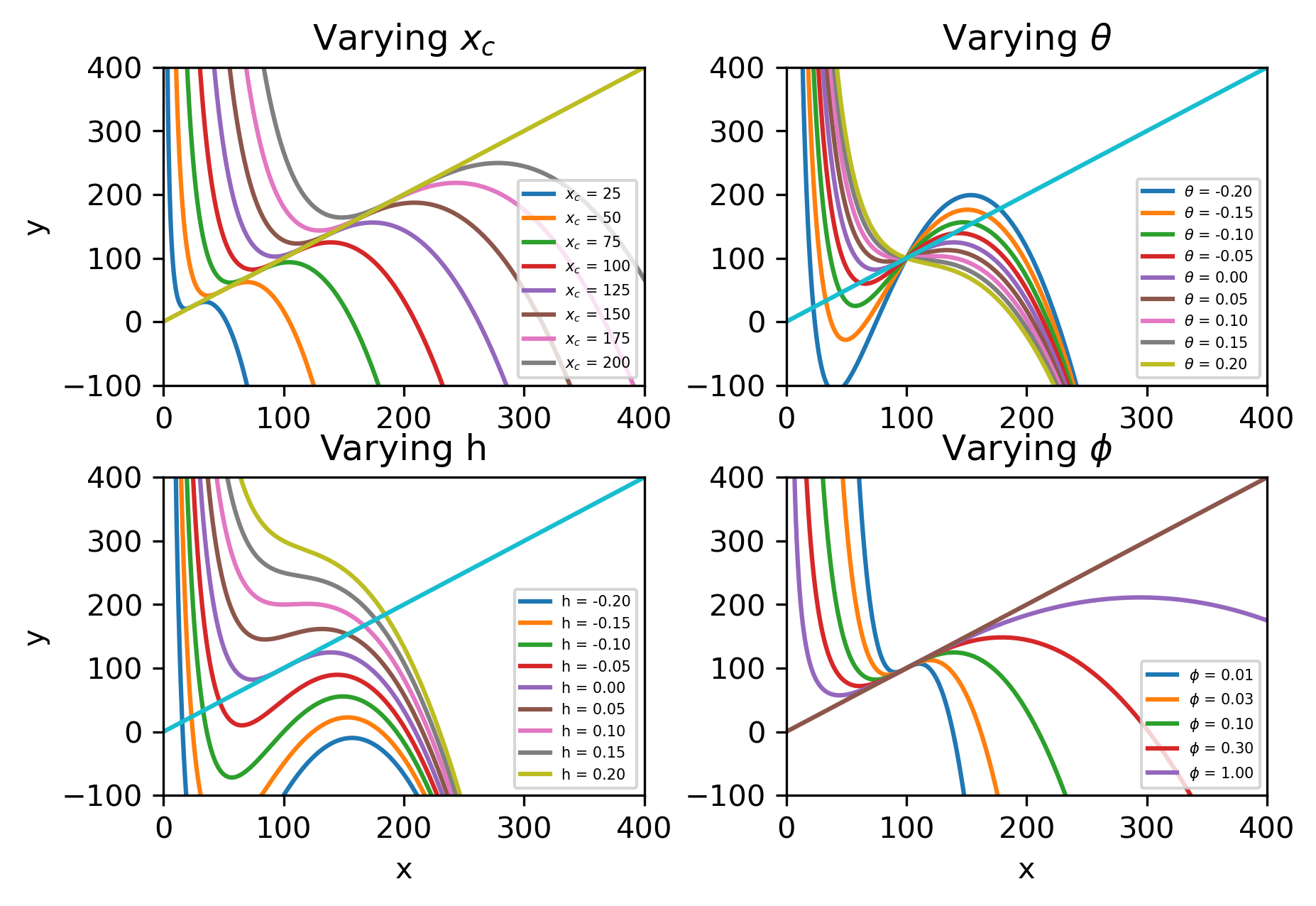}
    \caption{Parameter exploration of nullclines in x and y space.}
    \label{nullclines}
\end{figure}

\begin{figure}[h][ht]
    \centering
    \includegraphics[scale = .5]{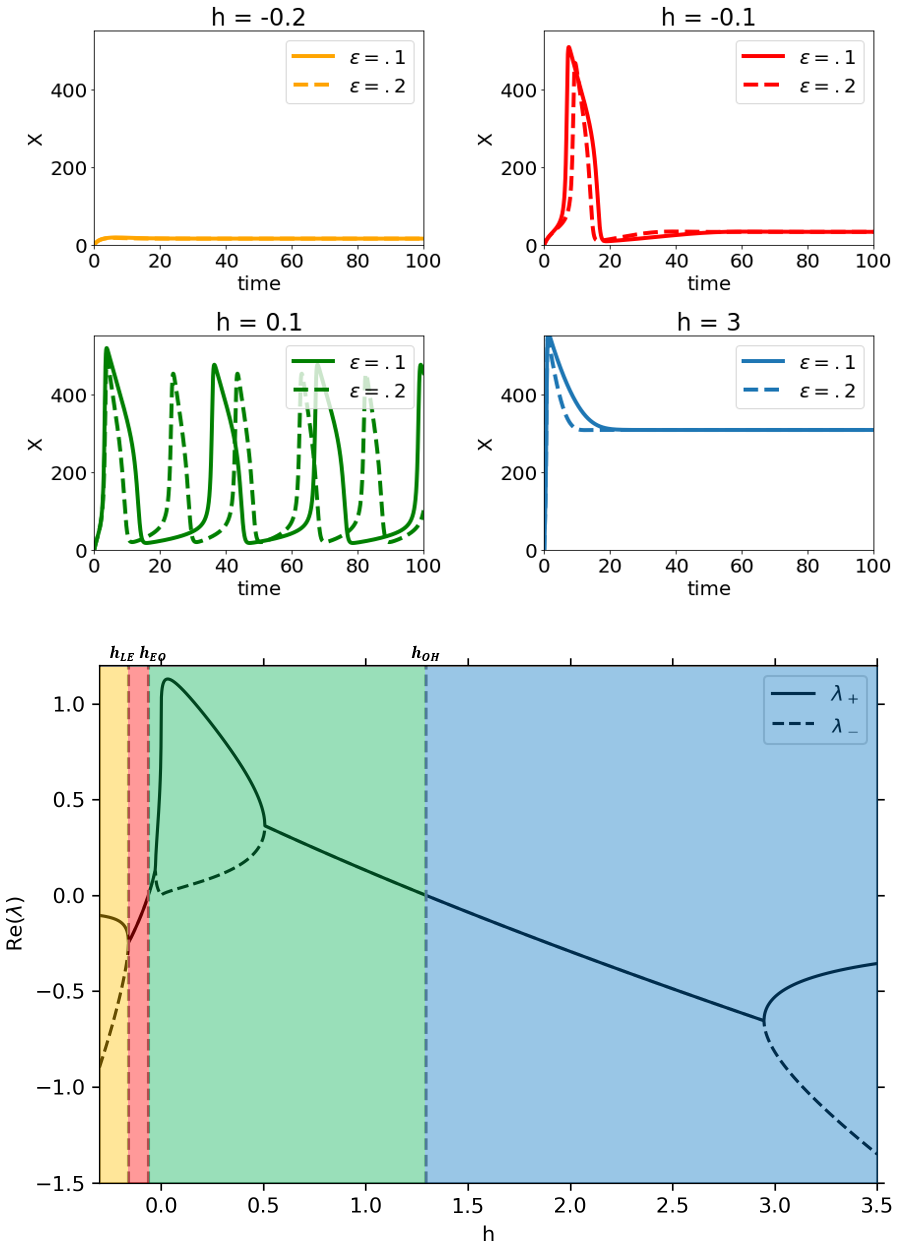}
    \caption{Top: The four possible distinct dynamical regimes plotted with different $\epsilon$ values. \quad
    Bottom: How the eigenvalues change with $h$. The colored regions match the color of the respective dynamical regime in the top portion of the plot.}
    \label{eigenvalues_vs_h}
\end{figure}

Working at the critical temperature ($\theta = 0$) and setting $k_6 = k_7 = \epsilon$ we solve 
for the stability of the fixed point. Equations \ref{m_dot2} and \ref{n_dot2} become
\begin{equation}
    \dot m = h - \frac{1}{3}m^3 - \phi (m+1)n
    \label{m_dot3}
\end{equation}
\begin{equation}
    \dot n = -\epsilon n - \dot m
    \label{n_dot3}
\end{equation}
Equation \ref{fixedpt3} gives the value of $m$ at the fixed point as 
\begin{equation}
    m = (3h)^{1/3}
\end{equation}
This gives the fixed point at $((3h)^{1/3},0)$ in $m$ and $n$ space. Linearizing around the fixed point 
gives eigenvalues of 
\begin{equation}
    \lambda _{\pm} = \frac{1}{2}(b - a - \epsilon \pm \sqrt{(a + \epsilon - b)^2 - 4a\epsilon}).
    \label{eigenvalues}
\end{equation}
where 
$$a = (3h)^{2/3}$$
$$b = \phi((3h)^{1/3}+1)$$
The bottom portion of Figure \ref{eigenvalues_vs_h} shows how the eigenvalues change with $h$.
\\

The system has four distinct dynamical regimes that can be seen in the top portion of Figure \ref{eigenvalues_vs_h}. 
In general, a fixed point is stable when both eigenvalues are negative 
or the real part of the eigenvalues is negative and unstable otherwise. 
For this system, the transitions from stable and unstable occur when

\begin{equation}
    0 = b - a - \epsilon
    \label{stability_condition}
\end{equation}
Substituting in the values of $a$ and $b$ into Eq.\ \ref{stability_condition} and solving for $h$ gives
\begin{equation}
    h = \frac{1}{24}[\phi \pm \sqrt{\phi^2 + 4(\phi - \epsilon)}]^3
    \label{stability_boundaries}
\end{equation}
Since the oscillatory regime is the only unstable regime, the two values of $h$ in Eq.\ \ref{stability_boundaries} 
give the transitions to and from it. This means that the larger value of $h$ gives
the boundary between the oscillatory and monostable high regimes and the smaller value of $h$ gives 
the boundary between the excitable and oscillatory regime.
\\

Real/complex transitions tell when the fixed point begins or stops some form of oscillations. These 
transitions occur when the discriminant vanishes

\begin{equation}
    (a + \epsilon - b)^2 = 4a\epsilon
    \label{real_complex}
\end{equation}
Equation \ref{real_complex} gives four different values for $h$.
\begin{equation}
    h = \frac{1}{3}[-\sqrt{\epsilon} + \frac{\phi}{2} \pm \frac{1}{2}\sqrt{4\phi - 4\phi \sqrt{\epsilon} + \phi^2}]^3
    \label{h_real_complex_low}
\end{equation}
\begin{equation}
    h = \frac{1}{3}[\sqrt{\epsilon} + \frac{\phi}{2} \pm \frac{1}{2}\sqrt{4\phi + 4\phi \sqrt{\epsilon} + \phi^2}]^3
    \label{h_real_complex_high}
\end{equation}
Because the excitable regime is stable low, the smallest value of $h$ in Equations \ref{h_real_complex_low} and 
\ref{h_real_complex_high} corresponds to the boundary between the monostable low and excitable regimes. 
To recap, the values of $h$ between the dynamical regimes are:

The transition between the monostable low and excitable regime: 
\begin{equation}
    h_{LE} = \frac{1}{3}[-\sqrt{\epsilon} + \frac{\phi}{2} - \frac{1}{2}\sqrt{4\phi - 4\phi \sqrt{\epsilon} + \phi^2}]^3
\end{equation}

The transition between the excitable and oscillatory: 
\begin{equation}
    h_{EO} = \frac{1}{24}[\phi - \sqrt{\phi^2 + 4(\phi - \epsilon)}]^3
\end{equation}

The transition between the oscillatory and monostable high regime: 
\begin{equation}
    h_{OH} = \frac{1}{24}[\phi + \sqrt{\phi^2 + 4(\phi - \epsilon)}]^3
\end{equation}

\subsection*{b. Granger Causality and Peak Order}

With the five original reactions and the Gillespie Stochastic Simulation Algorithm (GSSA), noisy molecular 
dynamics can be simulated. From these noisy trajectories, Granger causality analysis can be done. 
With a few cells, this poses no issue, but scaling up to the experimental size of $\approx 300$ cells is 
extremely computationally expensive. To keep run times realistic and the model as simple as possible, we use 
peak order as a proxy for Granger causality.
Intuitively, Granger causality determines whether one time series can help predict another time series. Given 
our system of exciting cells, it is not hard to imagine that if two neighboring cells both excite, one slightly 
before the other, that the faster cell's time series could be used to predict the slower's.

\begin{figure}[ht]
    \centering
    \includegraphics[scale = .75]{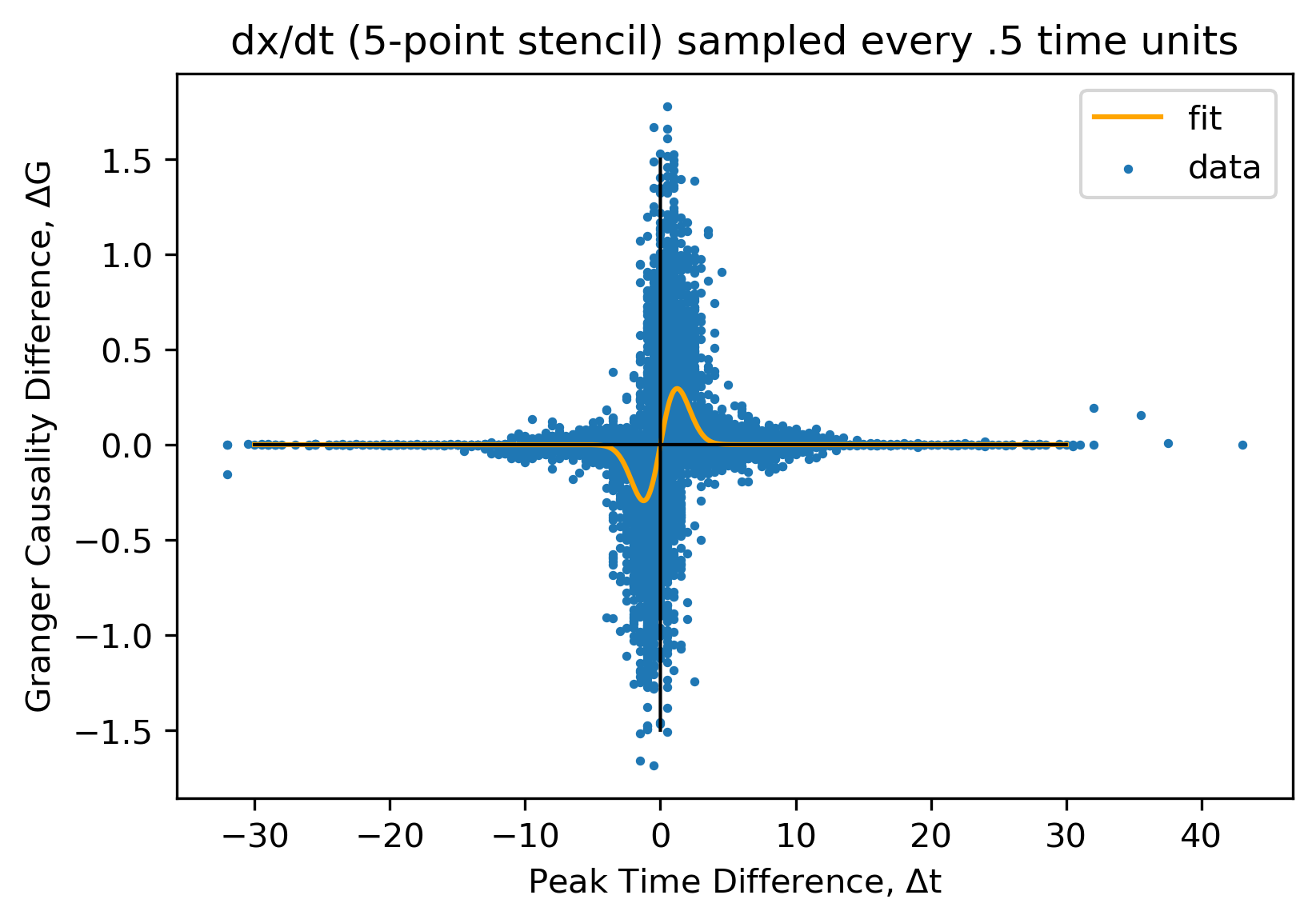}
    \caption{Fit line for thousands of simulations}
    \label{delta_G_vs_delta_t_raw}
\end{figure}

\begin{figure}[ht]
    \centering
    \includegraphics[scale = .75]{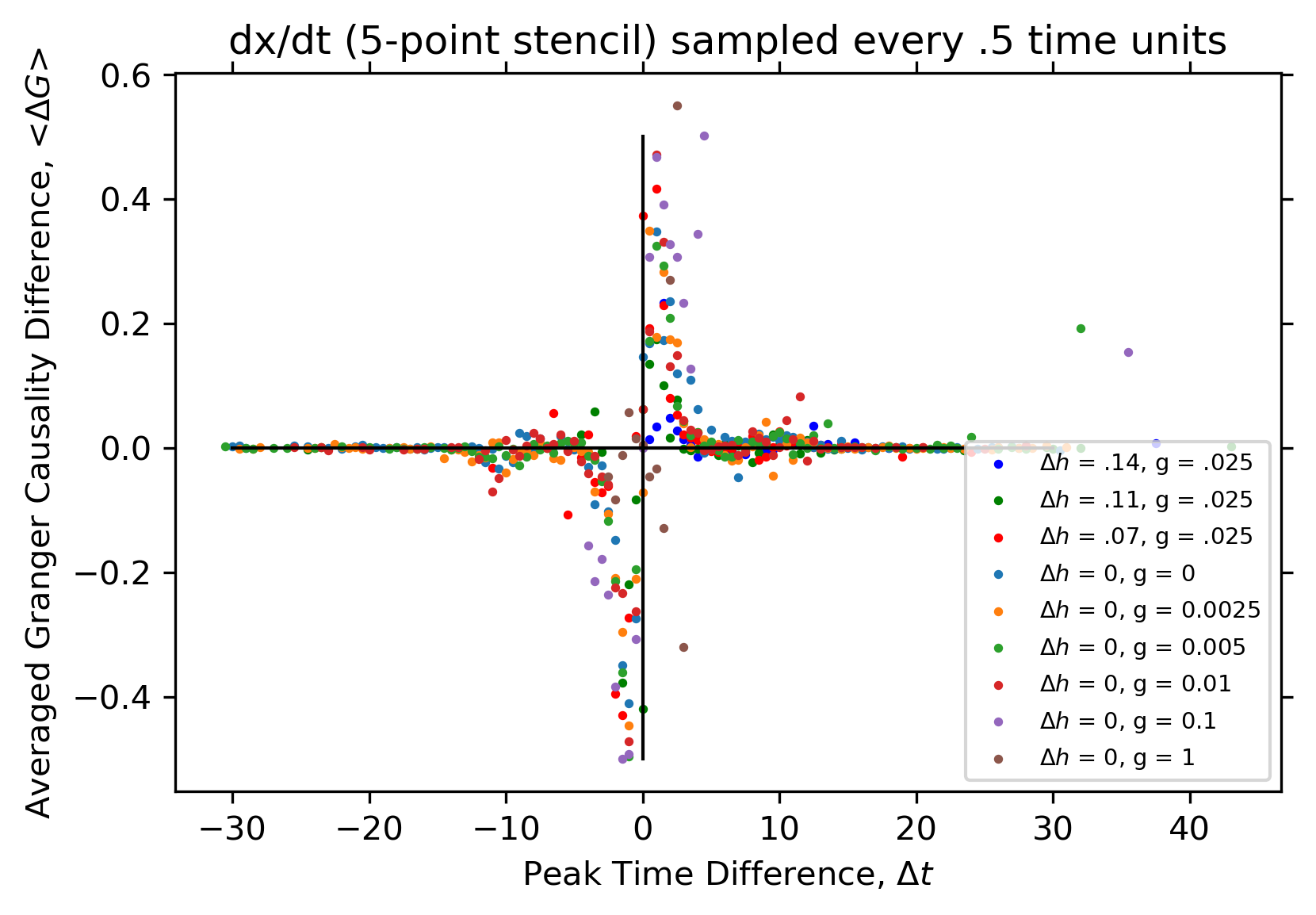}
    \caption{Data averaged in each time bin}
    \label{delta_G_vs_delta_t_average}
\end{figure}

To justify this proxy we ran thousands of GSSA simulations involving only two coupled cells that can exchange 
X molecules. The simulations involve different coupling strengths as well was different $h$ values for the cells. 
With the two simulated time series we calculated a differenced Granger metric of
\begin{equation}
    \Delta G = GC_{1 \rightarrow 2} - GC_{2 \rightarrow 1}.
    \label{delta_G}
\end{equation}
Where $GC_{1 \rightarrow 2}$ and $GC_{2 \rightarrow 1}$ are defined in subsection (b) of S2. To ensure that 
both time series were stationary, we first took a five-point stencil of both. The peak time difference, $\Delta t$,
is defined to be the time at which cell $2$ peaks minus the time at which cell $1$ peaks. Note that if 
cell 1 is Granger-causing cell 2 (positive $\Delta t$), $\Delta G$ should be positive and if cell 2 is Granger-causing cell 1 (negative $\Delta t$), 
$\Delta G$ should be negative. Figure \ref{delta_G_vs_delta_t_raw} shows a line fit to thousands of separate 
simulations. Figure \ref{delta_G_vs_delta_t_average} shows $\Delta G$ averaged over trials.
From these two plots it can be seen that on average, Granger causality and peak order coincide for a specific 
peak time difference window. We find this time window to be $[.5,2.5]$. Running simulations with hundreds of cells
is much more feasible when it is done with deterministic trajectories. If two neighboring cells in the lattice 
both excite and their peaks are within the chosen time window, an edge is counted between the two cells. 

\end{document}